\documentclass[11pt, letter]{article}
\usepackage{jheppub}

\usepackage[english]{babel}
\usepackage{blindtext}\blindmathtrue
\usepackage{avant} 
\usepackage[dvipsnames]{xcolor}
\usepackage{amsmath,epsf,amssymb,mathtools,latexsym,amsthm,setspace,array,pifont,amsfonts,dsfont,cancel,braket}
\usepackage{mathrsfs}
\usepackage{graphicx}
\usepackage{float}
\usepackage{tikz} 
\usepackage{tikz-feynman, contour}
\tikzfeynmanset{compat=1.1.0}
\usetikzlibrary{shapes,arrows,positioning,automata,backgrounds,calc,er,patterns}
\makeatletter
\newlength\CoolS@sizex
\newlength\CoolS@sizey
\newcommand*\CoolS@inner{%
\begin{tikzpicture}[baseline=0.04\CoolS@sizey]%
\foreach \x in {0, 1, ..., 5} \foreach \y in {0, 1, ..., 10}
\coordinate (c\x\y) at (\x *0.12*\CoolS@sizex, \y *0.107*\CoolS@sizey);
\draw [line width=\Cool@stroke] (c28)--(c26)--(c44)--(c42)--(c20)--(c02)--(c04)--(c15);
\draw [line width=\Cool@stroke] (c22)--(c24)--(c06)--(c08)--(c210)--(c48)--(c46)--(c35);
\end{tikzpicture}}
\usepackage[disable]{todonotes}

\graphicspath{{Pictures/}} 
\newcommand{\nn}{\nonumber}
\newcommand{\sd}{\mathrm{d}}
\newcommand{\pd}{\partial}

\newcommand{\bb}[1]{\mathbb{#1}}
\newcommand{\cl}[1]{\mathcal{#1}}

\def\prd{\ref@{Phys.~Rev.~D}}        
\newcommand{\Tr}[1]{\text{Tr}\left(#1\right)}
\newcommand{\td}[1]{
\if\notesOn1
\todo[inline]{#1}
\fi
}

\def\notesOn{1}
\tikzset{
	graviton/.style={
		double,
		decoration={snake, aspect=0.75, mirror, segment length=1.5mm},
		decorate
	}
}

\tikzfeynmanset{
	bigblob/.style={
		shape=circle,
		draw=blue,
		fill=red}
}

\renewcommand{\Re}{\operatorname{Re}}
\renewcommand{\Im}{\operatorname{Im}}

\def\mM{\mathcal{M}}

\def\mM{\mathcal{M}}

\def\mQ{\mathcal{Q}}

\def\mI{\mathcal{I}}

\def\mb{\bar{m}}

\def\pa{\partial}

\renewcommand{\vector}{\mathbf}
\renewcommand{\[}{\begin{equation}\begin{aligned}}
\renewcommand{\]}{\end{aligned}\end{equation}}
\def\rootKerr{$\sqrt{\textrm{Kerr}}$}

\def\sdyon{\textrm{spinning-dyon}}
\def\g5{\gamma_5}

\def\Qt{\tilde{Q}}

\def\Jt{\tilde{J}}

\def\Ct{\tilde{C}}

\def\b[#1]{\mathbf{#1}}
\def\bb[#1]{\overline{\mathbf{#1}}}
\def\bs[#1,#2]{\mathbf{#1}_{#2}}
\def\bbs[#1,#2]{\overline{\mathbf{#1}}_{#2}}

\def\s2{\sigma_2}
\def\ep{\epsilon}

\def\ketd[#1]{\ket{#1}_{\text{dressed}}}
\def\brad[#1]{\bra{#1}_{\text{dressed}}}
\def\ketas[#1]{\ket{#1}_{\text{Asymptotic}}}
\def\braas[#1]{\bra{#1}_{\text{Asymptotic}}}

\def\bn{\mathbf n}
\def\bae{{\boldsymbol{\mu}}_\textrm{elec}}
\def\bz{\mathbf z}
\def\ba{\mathbf a}
\def\bF{\mathbf F}
\def\bp{\mathbf p}
\def\bA{\mathbf A}
\def\bB{\mathbf B}
\def\bD{\boldsymbol{\nabla} }
\def\br{\mathbf r}
\def\bE{\mathbf E}
\def\bA{\mathbf A}
\def\bv{\mathbf v}
\def\dd{\hat{\mathrm{d}}}

\def\del{\hat \delta}
\def\eq{\begin{equation}}
\def\eqe{\end{equation}}
\def\eqa{\begin{eqnarray}}
\def\eqae{\end{eqnarray}}

\title{
Amplitudes from Coulomb to Kerr-Taub-NUT}
\author[1,2] {William T. Emond,}
\author[3,4]{Yu-tin Huang,}
\author[5]{Uri Kol,}
\author[6,7]{Nathan Moynihan}
\author[7]{and Donal O'Connell}

\affiliation[1]{School of Physics and Astronomy, University of Nottingham,
University Park, Nottingham NG7 2RD, United Kingdom}
\affiliation[2]{CEICO, Institute of Physics of the Czech Academy of Sciences, Na Slovance 1999/2, 182 21 Prague 8, Czech Republic}
\affiliation[3]{Department of Physics and Astronomy, National Taiwan University, Taipei 10617, Taiwan}
\affiliation[4]{Physics Division, National Center for Theoretical Sciences, National Tsing-Hua University, No.101, Section 2, Kuang-Fu Road, Hsinchu, Taiwan}
\affiliation[5]{ Center for Cosmology and Particle Physics, Department of Physics, New York University, 726 Broadway, New York, NY 10003, USA}
\affiliation[6]{The Laboratory for Quantum Gravity \& Strings, Department of Mathematics \& Applied Mathematics, University Of Cape Town}
\affiliation[7]{Higgs Centre for Theoretical Physics, School of Physics and Astronomy, The University of Edinburgh, EH9 3FD, Scotland}
\emailAdd{william.emond@nottingham.ac.uk}
\emailAdd{yutinyt@gmail.com}
\emailAdd{urikol@gmail.com}
\emailAdd{nathantmoynihan@gmail.com}
\emailAdd{donal@staffmail.ed.ac.uk}
\abstract{
Electric-magnetic duality, the Newman-Janis shift, and the double copy all act by elementary operations on three-point amplitudes. At the same time, they generate a network of interesting classical solutions spanning from the Coulomb charge via the dyon to the Kerr-Taub-NUT spacetime. We identify the amplitudes corresponding to each of these solutions, working to all orders in spin, but to leading perturbative order. We confirm
that the amplitudes double-copy when the solutions are related by the classical double copy.
Along the way we show that the Kerr-Taub-NUT solution corresponds to a gravitational electric-magnetic duality rotation acting on the 
Kerr solution, again to all orders in spin, and demonstrate that the asymptotic charges also transform simply under our operations.

}
\begin{document}

\maketitle

\section{Introduction}

Relations between classical solutions of gauge theory and gravity have been a subject of intense research for many decades.
We can classify these relations into two broad groups: maps between solutions of different theories, and maps between different solutions of the same theory. 
For instance, the double copy~\cite{Kawai:1985xq,Bern:2008qj,Bern:2010ue,Bern:2010yg} provides a concrete relation between gravitational solutions and those of Yang-Mills theory, so it is an
example of the first group of maps.
Meanwhile, examples of the second kind of maps are (Montonen-Olive) electric-magnetic duality, which rotates solutions in the same theory into each other, and the Newman-Janis shift~\cite{Newman:1965tw}, which generates rotating solutions from static ones in both electromagnetism (EM) and general relativity (GR).
We can view all of these maps as operations acting on classical solutions, thereby generating other solutions.
Our main purpose in this paper is to study the interplay between these three examples: the double copy, electric-magnetic duality, and the Newman-Janis shift.

In general, these operations could be technically involved and take complicated forms.
However, their manifestation on certain observables may be simple.
The key idea we explore in this work is that each of our three operations acts very simply on a scattering amplitude which describes
the leading order interaction between a classical background and a probe particle.
By following the action of the operations on solutions, and at the same time on amplitudes, 
we are able to associate a specific amplitude to each background. For example, we determine the three-point amplitude 
associated with the Kerr-Taub-NUT spacetime: a very exotic
object from the traditional perspective on scattering amplitudes. Astonishingly, the amplitude associated with Kerr-Taub-NUT is a very
simple object, closely connected to the amplitude for a photon coupling to a point electric charge.

Our work will be restricted to the context of linearised gravity. 
The gravitational version of electric-magnetic duality is best understood in linear theory, 
and in any case the comparisons we make between classical solutions and scattering
amplitudes are already non-trivial at linear order.
It is possible to perform more detailed comparisons between amplitudes and backgrounds at higher
orders, but we leave this for future work.
The linearised gravity approximation we have in mind (also known as the post-Minkowskian, or PM, approximation) corresponds to expanding full GR results to linear order in masses. In spite of this expansion, we will work to all orders in certain other parameters, notably the spin parameter $a$ of the Kerr metric. While the exact Kerr solution contains a naked singularity when $a$ is larger than the mass, there is no a priori reason to truncate the expansion in $a$ from the point of view of linearised gravity. 
Meanwhile recent progress in scattering amplitudes has shown that it is very natural to work to all orders in $a$. We also note 
the compelling simplicity of the results of Vines~\cite{Vines:2017hyw} on the linearised impulse in a Kerr background, to all orders in spin.

Combining our three operations will allow us to relate an interesting class of different solutions and their associated scattering amplitudes.
As we shall see, all three operations commute with one another. Let's begin with an overview of each operation separately.

The double copy is a relationship between scattering amplitudes
in a pair of theories. The original, and most well-studied, double copy\footnote{The double copy has recently been reviewed in detail~\cite{Bern:2019prr}.} expresses amplitudes in gravity in terms of amplitudes in
Yang-Mills theory. Although rooted in quantum field theory, amplitudes describe the time evolution from the far
past to the far future in any quantum theory in an asymptotically Minkowski spacetime, including in the classical approximation~\cite{Kosower:2018adc,Maybee:2019jus,delaCruz:2020bbn}. Therefore the double copy is present in at least certain 
aspects of classical gravity. In some cases this extends to an all-order relation between classical solutions of Yang-Mills theory
and gravity, a relation classed the classical double copy~\cite{Monteiro:2014cda,Luna:2015paa,Luna:2016due,Goldberger:2016iau,Lee:2018gxc,Luna:2018dpt,Adamo:2018mpq,CarrilloGonzalez:2019gof,Cho:2019ype,Carrillo-Gonzalez:2019aao,Moynihan:2019bor,Bah:2019sda,Huang:2019cja,Alawadhi:2019urr,Borsten:2019prq,Kim:2019jwm,Banerjee:2019saj,Bahjat-Abbas:2020cyb,Moynihan:2020gxj,Adamo:2020syc,Alfonsi:2020lub,Luna:2020adi,Keeler:2020rcv,Elor:2020nqe,Cristofoli:2020hnk,Alawadhi:2020jrv,Casali:2020vuy,Adamo:2020qru,Easson:2020esh,Chacon:2020fmr,Godazgar:2020zbv}. 
For example, a Coulomb charge double-copies to the Schwarzschild solution~\cite{Monteiro:2014cda}.

The second operation of interest to us is electric-magnetic duality. This is the rotation of electric and magnetic 
fields into one another:
\[\label{EMduality}
\vector E' &= +\cos \theta \, \vector E - \sin \theta \, \vector B\, , \\
\vector B' &= +\sin \theta \, \vector E + \cos \theta \, \vector B\, .
\]
By the equations of motion, the duality also rotates electric charges $q$ and magnetic charges $g$ so that
\[
q' &= +\cos \theta \, q - \sin \theta \, g\, , \\
g' &= +\sin \theta \, q + \cos \theta \, g\, .
\label{eq:chargeDuality}
\]
In particular, electric-magnetic duality rotates a Coulomb charge into a dyon. In terms of the covariant field strength $F_{\mu\nu}$ and its dual 
$\tilde F_{\mu\nu} = \epsilon_{\mu\nu\rho\sigma} F^{\rho\sigma}/2$, this duality is
\[
F'_{\mu\nu} &= +\cos \theta \, F_{\mu\nu} + \sin \theta \, \tilde F_{\mu\nu}\, , \\
\tilde F'_{\mu\nu} &= -\sin \theta \, F_{\mu\nu} + \cos \theta \, \tilde F_{\mu\nu}\, .
\]
(We provide details of our conventions in appendix~\ref{sec:conventions}.)

The combined operation of electric-magnetic duality together with the double copy gives rise to a gravitational duality. This gravitational electric-magnetic duality was studied in \cite{Huang:2019cja} in terms of asymptotic charges, but it can also be realized using curvature invariants.
Writing the linearized Riemann tensor as $R_{\mu\nu\rho\sigma}$, one can define a dual Riemann tensor
\[
\tilde R_{\mu\nu\rho\sigma} \equiv \frac 12 \epsilon_{\mu\nu\alpha\beta} R^{\alpha\beta}{}_{\rho\sigma} \,. 
\label{dual}
\]
Then the gravitational electric-magnetic duality is defined by the rotation
\[\label{duality}
R_{\mu\nu\rho\sigma}' &= +\cos \theta \, R_{\mu\nu\rho\sigma} + \sin \theta \, \tilde R_{\mu\nu\rho\sigma}\, , \\
\tilde R_{\mu\nu\rho\sigma}' &= -\sin \theta \, R_{\mu\nu\rho\sigma} + \cos \theta \, \tilde R_{\mu\nu\rho\sigma}\, .
\]
At least at the linearized level, such a duality operation is not new and was described well before the advent of the double copy (see for example \cite{Stephani:2003tm,Henneaux:2004jw,Bunster:2006rt,Argurio:2009xr}).
Here we will show that the electric-magnetic duality \eqref{EMduality} is related by the double copy to gravitational duality as defined in \eqref{duality}.

As a concrete example, the action of the gravitational electric-magnetic duality on the linearized Schwarzschild metric produces the linearized Taub-NUT metric \cite{Henneaux:2004jw,Bunster:2006rt,Argurio:2009xr}. The double copy further maps the Taub-NUT metric and the dyon solution into each other~\cite{Luna:2015paa}.
In particular, the mass aspect and the NUT parameter are mapped into the electric and magnetic charges, respectively.
We will refer to both the standard electric-magnetic duality and its gravitational counterpart simply as ``duality'' below.

The third (and final) operation of interest to us is the Newman-Janis shift~\cite{Newman:1965tw}. This shift can be viewed as a translation, though with an imaginary
parameter. Acting on the Schwarzschild solution, this shift leads to the Kerr metric.
The Newman-Janis shift was recently understood in terms of a remarkable exponentiation of spin effects at the level of three-particle  amplitudes~\cite{Arkani-Hamed:2019ymq}.
One can also apply the Newman-Janis shift in an electromagnetic context. For example, it relates the Coulomb charge to a
solution of the Maxwell equations describing a spinning charge configuration which was named \rootKerr ~in \cite{Arkani-Hamed:2019ymq}.

\begin{figure}[t]
\begin{center}
\includegraphics[width=1\textwidth]{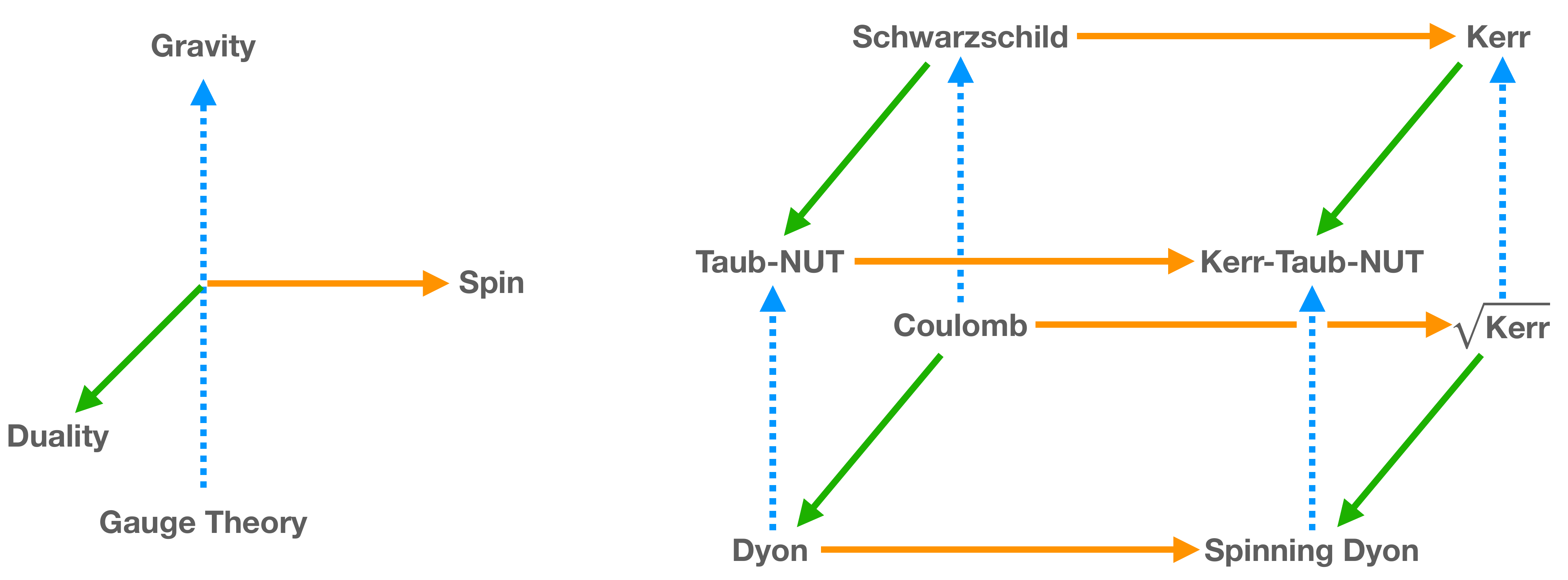}
\end{center}
\caption{Classical solutions related by the double copy (dashed blue arrow), duality (green arrow) and the Newman-Janis shift (orange arrow).
The double copy provides a map between solutions of different theories.
Duality and the Newman-Janis shift, on the other hand, map different solutions within the same theory and depend on continuous parameters.
}
\label{fig:solutionCube}
\end{figure}

As we have already seen, the three operations of our central interest relate a number of interesting solutions to one another.
In this article, we will address the full set of solutions related by these three operations acting one after another.
Indeed, a generalization of the Newman-Janis shift was studied by Talbot \cite{Talbot:1969bpa} and provides a wider set of maps between different solutions.
In particular, Talbot found a complex shift that maps the Schwarzschild metric into the Kerr-Taub-NUT spacetime.
We show that this complex shift, at the linearized level, can be viewed as a combination of duality and the Newman-Janis shift.
We will explore its physical origin and also study its single copy counterpart in gauge theory.
Taken together, our operations generate a network of eight different solutions which naturally has the structure of the cube shown in figure~\ref{fig:solutionCube}.

Insight into the physical origin of the complex shift can be gained by studying the asymptotic charges that characterize this class of solutions.
The group of asymptotic charges close an algebra that describe the asymptotic symmetries of spacetime.
This symmetry group is comprised of Bondi-Metzner-Sachs (BMS) transformations \cite{Bondi:1962px,Sachs:1962wk}, which include Lorentz transformation and supertranslations, as well as the recently discovered dual supertranslation symmetry \cite{Godazgar:2018qpq,Godazgar:2018dvh,Kol:2019nkc}.
We can then understand the action of the operations discussed above on the scattering amplitude as a manifestation of the asymptotic symmetry group.
In particular, the duality operation corresponds to a mixing of standard and dual supertranslation charges \cite{Huang:2019cja}.
We will see that the combined operation of duality and the Newman-Janis shift involves subleading BMS and dual BMS charges.

The vertical axis in figure~\ref{fig:solutionCube} represents the classical double copy. In spite of considerable recent work on the classical
double copy, it remains the case that the traditional double copy for scattering amplitudes is much better understood. In this paper
we are able to make a direct connection between both avatars of the double copy by directly identifying the amplitudes associated with
each of the eight solutions in the figure. We then check directly that the solutions we claim are related by the double copy behave as 
expected. We do so by computing a physical observable in two ways: firstly by using the equations of motion with the known
expression for the solutions, and secondly using amplitudes based on the methods of~\cite{Kosower:2018adc}. We verify agreement
between both approaches to assure ourselves that we have a consistent picture.

We emphasize again that while our results rely on the PM approximation, the spin parameter is left finite and unconstrained throughout this work.
That should be contrasted against most treatments of the Kerr-Taub-NUT spacetime in the literature, which consider different limits of the spin.

The paper is organized as follows.
We begin our discussion in section~\ref{sec:duality} by explaining in more detail how our three operations build the cube in figure~\ref{fig:solutionCube}. We dwell on duality in the gravitational context which is perhaps more unfamiliar, 
explaining the relationship between duality as we described it above and the asymptotic symmetry group of spacetime. We define
what we mean by a spinning dyon, and also show that the linearized Schwarzschild and Kerr-Taub-NUT metrics are related by the 
combined transformation of duality and the Newman-Janis shift to all orders in spin. 
In section \ref{sec:impulse} we compute the impulse imparted to a probe particle moving in the classical background using the equations of motion.
In section \ref{sec:amplitudes} we study how duality and the Newman-Janis shift act on certain massive three-particle amplitudes in gauge theory and in gravity. We then use the amplitude description to compute the impulse a second time, thereby verifying the double copy between
the classical solutions at linear order.
Throughout this paper we will be using the mostly plus metric signature $(-+++)$.

\section{A Network of Solutions}\label{sec:duality}

Each of the axes in figure~\ref{fig:solutionCube} corresponds to one of our operations. Let's now discuss how the operations take us
from one solution to the next, beginning with the simplest solution: the Coulomb charge.

As we mentioned in the introduction, it is an elementary fact that electric-magnetic duality relates the point Coulomb charge to the dyon.
It is less well-known that the Newman-Janis shift also acts in an interesting way on the Coulomb charge, though this action was also
discussed in Newman and Janis's original paper~\cite{Newman:1965tw}. We define the result of this action to be the \rootKerr~solution.

Restricting to the lower four (electromagnetic) solutions, it remains to discuss the spinning dyon. This deserves more than a few words, 
so we postpone the discussion to section~\ref{sec:spDyon} below.

The upper set of backgrounds in the figure are solutions to the Einstein equation. The Schwarzschild solution is, of course, 
the simplest of these. The
Newman-Janis shift was designed as a method to recover the Kerr metric from Schwarzschild. Duality in the gravitational context is less
well understood than in the electromagnetic case, so again it is worth discussing the duality relation in the gravitational context in
more detail. We do so below in section~\ref{sec:KTN} focusing especially on the Kerr-Taub-NUT case which is the most sophisticated 
example we consider.

Before discussing the solutions in more detail, it will be useful for us to review the Newman-Penrose formalism. This elegant formalism
allows us to demonstrate how duality as we described it in equation \eqref{duality} induces a transformation on asymptotic charges which 
was helpful for understanding Taub-NUT in~\cite{Huang:2019cja}.

\subsection{Newman-Penrose and Asymptotic Charges}

We introduce the Newman-Penrose (NP) formalism~\cite{Newman:1968uj} by choosing a null tetrad 
\[
e^{\mu}_a=\{ \ell^{\mu},n^{\mu},m^{\mu},\mb ^{\mu} \} \,,
\]
with
the property that $\ell^\mu$ is the tangent vector along outgoing lightlike geodesics. In Minkowski space, we can choose $\ell$ and $n$ to
be real vectors with $\ell \cdot n = -1$, while $m$ and $\mb = m^*$ are complex null vectors satisfying $m \cdot \mb = 1$. With this basis, 
we may define Newman-Penrose scalars for both the
electromagnetic and gravitational fields. In the electromagnetic case, given a field strength $F^{\mu\nu}$, we define
\[
\Phi_0 &= F_{\mu\nu} \ell^\mu m^\nu \,, \\
\Phi_1 &= \frac{1}{2} F_{\mu\nu} (\ell^\mu n^\nu + \bar m^\mu m^\nu) \,,\\
\Phi_2 &= F_{\mu\nu} \bar m^\mu n^\nu \, .
\]
These three complex scalars encode the six degrees of freedom in the field strength. 

Similarly, in the gravitational case,
the ten degrees of freedom in the metric can be organized in terms of five complex scalars using the Weyl tensor $C_{\mu\nu\rho\sigma}$ as follows
\begin{equation}
\begin{aligned}
\Psi_0 &= C_{\mu\nu\rho\sigma} \ell^{\mu}m^{\nu}\ell^{\rho}m^{\sigma}
\,,\\
\Psi_1 &= C_{\mu\nu\rho\sigma} \ell^{\mu}n^{\nu}\ell^{\rho}m^{\sigma}
\,,\\
\Psi_2 &= C_{\mu\nu\rho\sigma} \mb^{\mu}n^{\nu}\ell^{\rho}m^{\sigma}
\,,\\
\Psi_3 &= C_{\mu\nu\rho\sigma} \mb^{\mu}n^{\nu}\ell^{\rho}n^{\sigma}
\,,\\
\Psi_4 &= C_{\mu\nu\rho\sigma} \mb^{\mu}n^{\nu}\mb^{\rho}n^{\sigma} \,.
\end{aligned}
\label{eq:peeling}
\end{equation}
Imposing a retarded boundary condition, the expansion of the complex Weyl scalars in large radius $r$ takes the ``peeling'' form~\cite{Newman:1961qr}
\begin{equation}
\begin{aligned}
\Psi_0 &= \Psi_0^0 \frac{1}{r^5}+\Psi_0^1 \frac{1}{r^4} + \cdots , \\
\Psi_1 &= \Psi_1^0 \frac{1}{r^4}+\Psi_1^1 \frac{1}{r^5} + \cdots , \\
\Psi_2 &= \Psi_2^0 \frac{1}{r^3}+\Psi_2^1 \frac{1}{r^4} + \cdots , \\
\Psi_3 &= \Psi_3^0 \frac{1}{r^2}+\Psi_3^1 \frac{1}{r^3} + \cdots , \\
\Psi_4 &= \Psi_4^0 \frac{1}{r}+\Psi_4^1 \frac{1}{r^2} + \cdots ,
\end{aligned}
\end{equation}
Thus, $\Psi_0$ and $\Psi_1$ describe near-field components of the gravitational field that decay rapidly. $\Psi_2$ is dominant in the intermediate zone and describes the Coulomb components of the field. Finally, $\Psi_3$ and $\Psi_4$ describe the radiative components of the field that dominate at large distances.

It is sometimes useful to use a spinorial version of the Newman-Penrose formalism, which will be particularly familiar to readers who are acquainted
with the spinor-helicity formalism
in scattering amplitudes. Indeed the basic expressions in both of these formalisms are identical. Details of our spinor conventions are
given in appendix~\ref{sec:conventions}. 

Since the NP vector $\ell$ is null, we can 
find spinors $\ket{\ell}$ and $[\ell|$ such that $\ell = \ket{\ell} [\ell|$. Furthermore, $\ell^\mu$ is a real vector, so we choose the
anti-chiral spinor $[\ell|$ to be the complex conjugate of $\ket{\ell}$. Since the space of chiral spinors at a given spacetime point is two dimensional,
we complete the basis by picking another spinor $\ket{n}$ such that $\langle \ell n \rangle \neq 0$. More specifically we can choose $\ket{n}$ so that
$n = \ket{n} [n|$, where again $[n|$ is the complex conjugate of $\ket{n}$. 

The spinors $\ket{\ell}$ and $\ket{n}$ are a basis for the chiral spinors (and their conjugates are a basis for the anti-chiral spinors). We 
can therefore complete the basis of vectors by choosing $m = \ket{\ell}[n|$ with $\mb$ defined as the conjugate of $m$. In spinor-helicity terms, 
we are treating $\ell$ as a light-like momentum, $n$ as a gauge choice, and the pair $m$ and $\mb$ as polarisation vectors of definite helicity.

The spinorial equivalents of the $\Phi_i$ and $\Psi_i$ scalars are simply the spinorial forms of the curvature. To discuss spinors in curved space,
we first introduce a frame $e_a^\mu$ for $a = 0, \cdots, 3$ such that
\[
e_a^\mu g_{\mu\nu} e_b^\nu &= \eta_{ab} \,, \\
e_a^\mu \eta^{ab} e_b^\nu &= g^{\mu\nu} \,.
\]
We define the Maxwell spinor as
\[
\Phi_{\alpha\beta} =  F_{\mu\nu} \, e^\mu_a e^\nu_b \, \sigma^{ab}_{\alpha\beta} = F_{\mu\nu} \sigma^{\mu\nu}_{\alpha\beta} \,,
\label{eq:MaxwellSpinor}
\]
where 
\[
\sigma^{\mu\nu} = e_a^\mu e_b^\nu \sigma^{ab} = \frac14  e_a^\mu e_b^\nu \ \left( \sigma^a \tilde \sigma^b - \sigma^b \tilde \sigma^a \right) \,.
\]
Note that $\sigma^{ab}_{\alpha\beta}$ is symmetric in its spinor indices $\alpha$ and $\beta$. Consequently, 
$\Phi_{\alpha\beta}$ is a symmetric 
two-by-two matrix, containing three (complex) components. These are nothing but the NP scalars $\Phi_0$, $\Phi_1$ and $\Phi_2$. 

We similarly define the Weyl spinor $\Psi_{\alpha\beta\gamma\delta}$ as
\[
\Psi_{\alpha\beta\gamma\delta} = C_{\mu\nu\rho\sigma} \sigma^{\mu\nu}_{\alpha\beta} \sigma^{\rho\sigma}_{\gamma\delta} \,.
\]
The Weyl spinor is totally symmetric under its four spinor indices, and therefore contains precisely five components: the five $\Psi_i$ NP scalars.

The Newman-Penrose form is inherently chiral, treating fields of given helicity differently (equivalently, self-dual and anti-self-dual fields are treated
differently). It therefore diagonalises the operation of duality. To see this, it is worth noticing that the Maxwell spinor can be written in terms of the 
Pauli sigma matrices $\boldsymbol{\sigma}$ as
\[
\Phi_\alpha^\beta = (\vector{E} - i \vector{B}) \cdot \boldsymbol{\sigma}_\alpha^\beta \,,
\]
where $\vector{E}$ and $\vector{B}$ are the electric and magnetic fields, respectively. Thus under the duality~\eqref{EMduality}, we see that 
\[
\Phi' = e^{-i \theta} \Phi \,.
\label{eq:dualityOnMaxwellSpinor}
\]
As we will soon see, the gravitational case is identical.

Locally asymptotically flat spacetimes are invariant under the BMS group, which is composed of Lorentz transformations and supertranslations, 
as well as the recently discovered dual supertranslation symmetry \cite{Godazgar:2018qpq,Kol:2019nkc}. The standard and dual 
supertranslation charges are given by the leading order components of the gravitational analogue of the Coulomb field, integrated over the 
two sphere:
\begin{equation}
\begin{aligned}
T(f) &=  - \frac{1}{4 \pi G}\int _{S^2} \sd\Omega \, \sqrt{\gamma} \, f(\Omega)\,  \left[ \Re \Psi_2^0 (u,\Omega)\right]_{\mI^+_-},\\
\mM(f) &=  - \frac{1}{4 \pi G}\int _{S^2} \sd\Omega \, \sqrt{\gamma}   \, f(\Omega)\,  \left[ \Im \Psi_2^0 (u,\Omega)\right]_{\mI^+_-},\\
\end{aligned}
\end{equation}
respectively. Here $\Omega$ represent the two coordinates on the celestial sphere, $\gamma$ is the metric on unit sphere, $f(\Omega)$ is the transformation parameter, $u$ is the retarded null coordinate and $\mI^+_-$ is the lower boundary ($u=-\infty$) of future null infinity $\mI^+$.
One can then define a complex charge
\begin{equation}\label{leadBMS}
Q(f) \equiv T(f) + i \mM (f) = - \frac{1}{4 \pi G}\int _{S^2} \sd\Omega \, f(\Omega)\,  \left[  \Psi_2^0 (u,\Omega)\right]_{\mI^+_-}
\end{equation}
and a corresponding duality operation \cite{Huang:2019cja}
\begin{equation}\label{def2}
Q' (f) = e^{-i \theta} Q(f) .
\end{equation}
Equivalently, this duality rotates the standard and dual supertranslation charges into one another
\begin{equation}\label{ChargeDuality}
\begin{aligned}
T' (f) &= + \cos \theta \, T(f) + \sin \theta \, \mM (f) \,, \\
\mM ' (f) &= - \sin \theta \, T(f) + \cos \theta \, \mM (f) \,.
\end{aligned}
\end{equation}

It is interesting to write these charges in terms of gravitational analogues of electric and magnetic fields. These are simply a decomposition
of the Weyl tensor given by
\begin{equation}
\begin{aligned}
E_{\nu\sigma} &=  C_{\mu\nu\rho\sigma} \left(\ell^{\mu} + n^{\mu}\right)\left(\ell^{\rho} + n^{\rho}\right) , \\
B_{\nu\sigma} &=  \Ct_{\mu\nu\rho\sigma} \left(\ell^{\mu} + n^{\mu}\right)\left(\ell^{\rho} + n^{\rho}\right),
\end{aligned}
\end{equation}
where $\Ct_{\mu\nu\rho\sigma}$ is the dual of the Weyl tensor. Therefore the duality~\eqref{duality} for the curvature (and hence for the
Weyl curvature) is
\[
E_{\nu\sigma}' &= + \cos \theta \, E_{\nu\sigma} + \sin \theta \, B_{\nu\sigma}\,, \\
B_{\nu\sigma}' &= - \sin \theta \, E_{\nu\sigma} + \cos \theta \, B_{\nu\sigma} \,.
\label{eq:gravEBdual}
\]

Now, Newman and Penrose \cite{Newman:1968uj} showed that to leading order in the asymptotic expansion
\begin{equation}
\Psi_2 = Z_{\mu\nu} m^{\mu} \mb^{\nu}
\end{equation}
where
\begin{equation}
Z_{\mu\nu} = \frac{1}{2} E_{\mu\nu} + \frac{i}{2} B_{\mu\nu.}
\end{equation}
Thus, the duality of equation~\eqref{eq:gravEBdual} is simply a rephasing of the Newman-Penrose scalar
\[
\Psi_2' = e^{-i \theta} \Psi_2 \,.
\label{eq:dualityPsi2}
\]
But this is simply the duality~\eqref{def2} for the charges.

In reference~\cite{Huang:2019cja}, the duality of the charges was used to demonstrate that Taub-NUT is a duality rotation of Schwarzschild,
justifying the relation between the two show in figure~\ref{fig:solutionCube}.

\subsection{Spinning Dyons from Duality}
\label{sec:spDyon}

Now we return to figure~\ref{fig:solutionCube}, and we ask the question: what precise electromagnetic solution do 
we obtain by operating on the Coulomb charge with duality and the Newman-Janis shift?

The action of duality and the shift are understood separately. As we mentioned in the introduction, it is a basic fact that duality 
rotates a point electric charge into a dyon. Newman and Janis described the action of their shift on the Coulomb source 
in electromagnetism~\cite{Newman:1965tw},
resulting in a configuration of charge which has an associated angular momentum, and so a spin. It is a particular kind of
rotating disc of charge, and can be viewed as the electromagnetic field of the charged (Kerr-Newman) black hole.
More recently, it was understood~\cite{Arkani-Hamed:2019ymq} that this rotating disc solution is related by the double copy~\cite{Monteiro:2014cda} to
the Kerr solution, 
so the rotating disc of charge which was called \rootKerr~in reference~\cite{Arkani-Hamed:2019ymq}. We shall use this terminology 
below. The field strength of \rootKerr~is most conveniently discussed using the Maxwell 
spinor~\eqref{eq:MaxwellSpinor}.
We write the Maxwell spinor of a Coulomb charge at the origin as $\Phi_{\textrm{Coulomb}}(x)$.
With this notation, the \rootKerr~field strength spinor is
\[
\Phi_{\sqrt{\textrm{Kerr}}}(x) = \Phi_{\textrm{Coulomb}}(x+ia) \,,
\]
where $a^\mu$ is a vector parameterising the spin of the object\footnote{In the gravitational case of Kerr or Kerr-Taub-NUT, it is standard to choose $a^\mu$ to be in the $z$ direction.}.

It is straightforward to construct the Maxwell spinor of the spinning dyon, which we define to be the electromagnetic field configuration we obtain by
acting with duality on the field of \rootKerr. We have seen that the duality acts by rephasing~\eqref{eq:dualityOnMaxwellSpinor} the Maxwell 
spinor, so
\[
\Phi_{\sdyon}(x) &= e^{-i \theta} \Phi_{\sqrt{\textrm{Kerr}}}(x) \\
&= e^{-i \theta} \Phi_{\textrm{Coulomb}}(x+ia) \,.
\label{eq:PhiSpDyon}
\]
Notice that, like Kerr-Taub-NUT, the spinning dyon depends on three parameters: the electric charge, the magnetic charge, and the spin.

Returning once again to the class of theories shown in figure~\ref{fig:solutionCube}, we have now described each solution as well as
the links between them induced by duality and the Newman-Janis shift. Several of the double copy relations shown on the figure are
also standard: for relation between Coulomb and Schwarzschild~\cite{Monteiro:2014cda} is among the most basic examples of the double
copy, and is easily generalised to relate \rootKerr~and Kerr~\cite{Monteiro:2014cda}. The double copy relating a dyon and Taub-NUT was
described in terms of double Kerr-Schild solutions in~\cite{Luna:2015paa} and more recently investigated in terms of amplitudes and
the impulse~\cite{Huang:2019cja}. 
It remains for us to justify the double copy relation we claim holds between the spinning dyon and Kerr-Taub-NUT. In fact, the spinning dyon
was previously~\cite{Luna:2018dpt} proposed as the ``single'' copy of Kerr-Taub-NUT. We will demonstrate below that this proposal is
correct. First we show that gravitational duality relates Kerr-Taub-NUT to Kerr (to all orders in spin). Later, in section~\ref{sec:amplitudes}, we
will demonstrate that the standard double copy at the level of amplitudes is also consistent with this identification.

\subsection{Kerr-Taub-NUT from Duality}
\label{sec:KTN}

It remains for us to justify our claim that Kerr-Taub-NUT is related by duality to Kerr. Of course it is
natural to guess that Kerr-Taub-NUT must be the result of duality acting on Kerr. But, as we will see, it is not too hard to demonstrate the
result to all orders in spin.

The exact Kerr-Taub-NUT metric is given by
\begin{equation}\label{KTNmetric}
\sd s^2 =  -f \left(\sd t + \Omega \, \sd \phi  \right)^2
+\frac{\rho^2}{\Delta} \sd r \,\! ^2 +\rho^2 \left(\sd\theta ^2 + \sigma ^2 \sin ^2 \theta \, \sd\phi  ^2 \right) 
\end{equation}
where
\begin{equation}
\Omega = -2 \ell \cos \theta - (1-f^{-1}) a \sin ^2 \theta, \qquad    \Delta = r \,\! ^2 -2mr +a ^2 -\ell^2, \qquad    \sigma^2 =\frac{\Delta}{f \rho^2},
\end{equation}
and
\begin{equation}
f= 1- \frac{2mr +2 \ell (\ell -a \cos \theta)}{\rho^2}, \qquad \rho ^2 = r \,\! ^2 +(\ell -a \cos \theta)^2.
\end{equation}
This metric depends on three parameters: mass $m$, NUT charge $\ell$ and spin $a$. We work in units where all of these parameters have
dimension of length; later on, we will use the symbol $M$ for the mass (with dimension of mass) of particles.\footnote{The Taub-NUT solution and its rotating generalisation contain string singularities, which were recently discussed in~\cite{Huang:2019cja}. They will not be relevant below.}

The solution is of Petrov type D, so the only non-vanishing Newman-Penrose scalar is $\Psi_2$. In an appropriate frame, 
this scalar is~\cite{Kinnersley:1969zza}
\[
\Psi_2 = \frac{m+i \ell}{\left(r-i\ell -ia \cos \theta \right)^3} \,.
\label{eq:ktnPsi2}
\]
Some more familiar solutions can be obtained as special cases of Kerr-Taub-NUT. The Kerr metric is reproduced by setting the NUT parameter to zero; it describes a rotating black hole solution with angular momentum
\begin{equation}
J=am \,. 
\end{equation}
In a similar way, one can set the mass to zero; then the resulting solution describes a rotating NUT spacetime with dual angular momentum
\begin{equation}
\Jt = a \ell \,.
\end{equation}
We will soon show that the duality operation \eqref{duality} interchanges the roles of the angular momentum $J$ and its dual $\Jt$ (as well as interchanging the mass with the NUT parameter).

Talbot \cite{Talbot:1969bpa} found that the exact Kerr-Taub-NUT metric can be obtained from the Schwarzschild metric using a complex coordinate transformation given by
\begin{equation}
\begin{aligned}
t  &\longrightarrow  t  - i \ell + 2 i \ell \log \sin \theta \,, \\
r   &\longrightarrow r  - i \ell + i a \cos \theta \,.
\end{aligned}
\end{equation}
The complex coordinate technique is given by a non-rigorous algorithm whose physical origin is not fully understood. For a recent discussion we refer the reader to \cite{Huang:2019cja}. Here, working in linearised gravity, we interpret this complex shift as a combined operation of duality and the Newman-Janis shift, bearing in mind that each one of these operations was separately understood in \cite{Arkani-Hamed:2019ymq,Huang:2019cja}.

We work throughout to leading order in the linearised (or post-Minkowskian) approximation to general relativity. The explicit definition of
this limit we use is
\begin{equation}
\begin{aligned}
\frac{G M_{\rm ADM}}{r}=\frac{m}{r} &\ll 1\,,
\\
\frac{\ell }{r} &\ll 1\,.
\end{aligned}
\end{equation}
Notice that we impose no conditions on the spin $a$. In fact, the metric~\eqref{KTNmetric} is flat for any value of $a$ if we set both $m$ and
$\ell$ to zero. Thus the first approximation beyond Minkowski forces us to work to linear order in $m$ or $\ell$, or both. We will take $a$ to
be arbitrary for the rest of the paper, except where explicitly stated.

In this leading approximation, the Kerr-Taub-NUT metric takes the form
\begin{equation}
\label{eq:linKTNmetric}
\sd s^2_{\text{Kerr-Taub-NUT}} = \sd s^2_{\text{Flat}} + \sd s^2_{\text{Kerr}} + \sd s^2_{\text{rotating-NUT}} \,,
\end{equation}
where
\begin{equation}
\sd s^2_{\text{Flat}}  = -\sd t^2
+\frac{   r^2+a^2 \cos ^2\theta   }{r^2+ a^2}\sd r^2 
+ \left( r^2+ a^2 \cos ^2\theta \right)\sd\theta^2
+ \left( r^2 + a^2\right) \sin ^2\theta  \, \sd\phi ^2 
\end{equation}
is the flat metric in spheroidal coordinates,
\begin{equation}
\sd s^2_{\text{Kerr}}  =
2m r  \frac{1}{  r^2 +a^2 \cos ^2\theta }   \left(\sd t-a \sin ^2\theta   \, \sd\phi  \right)^2
+2m r   \frac{   r^2+a^2 \cos ^2\theta   }{(r^2+ a^2)^2}   \sd r^2 
\label{eq:linKerr}
\end{equation}
is the additional contribution due to the linearized Kerr metric and
\begin{equation}
\sd s^2_{\text{spinning-NUT}}=
-2 l \cos \theta 
\left( 
\frac{a \, \sd t^2
	-\left(r^2+a^2\right) (2 \sd t\, \sd\phi   -a \sin ^2\theta  \, \sd\phi ^2   )
}
{r^2+a^2 \cos ^2\theta }
+\frac{a }{r^2 + a^2}  \sd r^2 
+a  \, \sd\theta ^2 
\right)
\label{eq:linRotNUT}
\end{equation}
is the spinning NUT contribution. 

It will be useful for us to introduce a form of the curvature which extracts the linear dependence on mass $m$ and NUT charge $\ell$. So
we write the Kerr curvature tensor as
\[
\label{eq:ThetaKerrDef}
R^\textrm{Kerr}_{\mu\nu\rho\sigma} = m \, \Theta^\textrm{Kerr}_{\mu\nu\rho\sigma} \,,
\]
where $\Theta^\textrm{Kerr}$ is independent of $m$.
Similarly we may write the contribution of the spinning NUT to the total curvature of Kerr-Taub-NUT as
\[
\label{eq:ThetaSnutDef}
R^\textrm{spinning-NUT}_{\mu\nu\rho\sigma} = \ell \, \Theta^\textrm{spinning-NUT}_{\mu\nu\rho\sigma} \,.
\]
Thus the total linearised curvature of Kerr-Taub-NUT is
\[
R^\textrm{KTN}_{\mu\nu\rho\sigma} = m \, \Theta^\textrm{Kerr}_{\mu\nu\rho\sigma} + \ell \, \Theta^\textrm{spinning-NUT}_{\mu\nu\rho\sigma} \,.
\]

The Newman-Penrose form of the curvature~\eqref{eq:ktnPsi2} takes a particularly simple form in the linearised theory, which (up to an
irrelevant overall normalisation) is
\[
\Psi_2 = \frac{m+i \ell}{\left(r-ia \cos \theta \right)^3} \,.
\label{eq:linPsi2}
\]
It is straightforward now to take advantage of our understanding of the action~\eqref{eq:dualityPsi2} of duality on $\Psi_2$,
which immediately shows that duality acts on the mass and NUT charge as
\[
m' &= +\cos \theta \, m + \sin \theta \, \ell\, , \\
\ell' &= -\sin \theta \, m + \cos \theta \, \ell\, \,
\label{eq:massDuality}
\]
to all orders in the spin $a$. (Indeed the NUT charge is also referred to as the magnetic mass.) 
Since the Newman-Penrose scalars capture the full curvature of the spacetime, we conclude that duality rotates the linearised Kerr metric~\eqref{eq:linKerr} into the linearised rotating NUT metric~\eqref{eq:linRotNUT}. In other words,
\begin{equation}\label{dualityRelation}
\framebox[4.5cm][c]{$\displaystyle \tilde \Theta_{\mu\nu\rho\sigma}^{\text{Kerr}}=\Theta_{\mu\nu\rho\sigma}^{\text{spinning-NUT}}.$}
\end{equation}

It is possible to demonstrate the relationship \eqref{dualityRelation} directly at the level of the Riemann tensor rather than the Weyl spinor, at
the expense of lengthier
expressions. In appendix~\ref{app:duality} we describe one way to perform the calculation, and explicitly apply the method to the $a = 0$
case of equation~\eqref{dualityRelation}. We attach a mathematica notebook repeating the computation with $a \neq 0$.
Explicit expressions in the limit $a\ll 1$ with $J$ and $\Jt$ kept fixed can be found in \cite{Argurio:2009xr}.

The result \eqref{dualityRelation} implies that one can map the Schwarzschild and Kerr-Taub-NUT metrics into each other, in the leading PM approximation, using the combined operation of duality and the Newman-Janis shift. Referring back to our class of solutions, these
observations justify the links between the upper group of four solutions in figure~\ref{fig:solutionCube}. 

We will soon be interested in computing the impulse on a probe particle moving in the Kerr-Taub-NUT background. A direct computation
using the Kerr-Taub-NUT metric is non-trivial, so we will exploit duality to reduce the complexity. 
Using our
result~\eqref{dualityRelation}, we can conveniently write the linearised Kerr-Taub-NUT Riemann curvature as
\[
\label{eq:KTNcurvature}
R^\textrm{KTN}_{\mu\nu\rho\sigma} &= m \, \Theta^\textrm{Kerr}_{\mu\nu\rho\sigma} + \ell \, \Theta^\textrm{spinning-NUT}_{\mu\nu\rho\sigma} \\
&= \Re GM e^{-i \theta} \left( \Theta^\textrm{Kerr}_{\mu\nu\rho\sigma} + i  \tilde \Theta^\textrm{Kerr}_{\mu\nu\rho\sigma} \right) \,,
\]
where $m = GM \cos \theta$ and $\ell = GM \sin \theta$\,.

We finish our overview of duality in gravity with some comments on duality and asymptotic charges.
The duality relation between the angular momentum and the dual angular momentum has been studied in terms of Lorentz and dual Lorentz charges in \cite{Argurio:2009xr}. Here we would like to demonstrate that the Lorentz charges can be realized as subleading BMS charges that were studied in \cite{Godazgar:2018vmm}.
The first subleading complex BMS charge was derived in \cite{Godazgar:2018vmm} and it is given by
\begin{equation}\label{subleadingCcharge}
\mQ _ 1 (f) = - \frac{1}{8\pi G} \int \sd \Omega \, \sqrt{\gamma}\,  f(\Omega) \, \left(\Psi_2^1  -\bar{\eth} \Psi_1^0   \right).
\end{equation}
where $\bar{\eth}$ is a differential operator acting on $\Psi_1^0$, and the large $r$ (peeling) expansion coefficients $\Psi_i^j$ were defined in equation~\eqref{eq:peeling}. In our application $\Psi_1 = 0$ so we will not need the form of $\bar{\eth}$. By expanding the linearised Newman-Penrose scalar of equation~\eqref{eq:linPsi2}, we see that the two leading order coefficients are
\begin{equation}
\begin{aligned}
\Psi_2^0& = m + i \ell \,, \\
\Psi_2^1 &= 3 i a(m+ i \ell) \cos \theta \,.
\end{aligned}
\end{equation}
Recall that $\Psi_2^0$ determines the values of the leading BMS charges \eqref{leadBMS}, and as we have just seen, duality exchanges the roles of the mass $m$ and the NUT parameter $\ell$. The real and imaginary parts of the subleading BMS charge $\mQ _ 1$ take the following values
\begin{equation}
\begin{aligned}
\Re \mQ _ 1 (f) &= 
- \frac{3 \Jt}{8\pi G} \int \sd \Omega \,  f(\Omega)  \cos \theta, \\
\Im \mQ _ 1 (f) &= 
+ \frac{3 J}{8\pi G} \int \sd \Omega \,  f(\Omega)  \cos \theta.
\end{aligned}
\end{equation}
The result \eqref{dualityRelation} therefore implies that duality interchanges the roles of the real and imaginary parts of $\mQ_1$, or equivalently the angular momentum $J$ and the dual angular momentum $\Jt$.

\section{The Impulse}\label{sec:impulse}

In this section, we compute the impulse imparted to a probe scalar particle in the backgrounds of a spinning dyon and Kerr-Taub-NUT, at leading approximation using the equations of motion. The manifestation of the double copy structure is apparent in the impulse.
In addition, we show explicitly how the double copy structure relates the electromagnetic and the gravitational forces in those cases.

\subsection{Spinning Dyon}

In electromagnetic theory, we obtain the impulse by performing the total time integral of the Lorentz force. To get started, then, we need
to know the field strength tensor $F^{\mu\nu}(x)$ of a spinning dyon. First, let us define a little notation. We will take particle 1 to be the
probe, scattering off the heavy, static particle 2. At leading order, we may take the worldlines of both particles to be the straight lines
\[
\label{eq:LOtrajectory}
x_i(\tau) = b_i + u_i \tau \,. 
\]
We could choose one of the $b_i$ to be zero by choice of origin if we wished. The leading order proper velocities are $u_i$. Finally, we chose
the electric charge of our probe to be $e_1$.

We described how to determine the Maxwell spinor of the dyon in section~\ref{sec:spDyon}. The field strength in its usual form is then
simply
\[
F^{\mu\nu}_\textrm{spinning-dyon}(x)  = - \Re \Tr{\sigma^{\mu\nu} \Phi_\textrm{spinning-dyon}(x)} \,,
\]
where the spinning dyon's Maxwell spinor is given by equation~\eqref{eq:PhiSpDyon} in terms of the simple Coulomb Maxwell spinor
\[
\Phi_\textrm{Coulomb}(x) = 2 i e_2 \int \dd^4 q \, \del(q \cdot u_2) \, e^{iq \cdot (x - b_2)} \, \frac{q^\mu u_2^\nu \sigma_{\mu\nu}}{q^2} \,.
\]
With the help of the identity
\[
\Tr{ \sigma^{\mu\nu} \sigma^{\rho\sigma} } = - \frac12 \left( \eta^{\mu \rho} \eta^{\nu \sigma} - \eta^{\mu \sigma} \eta^{\nu \rho} + i \epsilon^{\mu\nu\rho\sigma} \right) \,,
\]
it is then easy to show that the field strength of the spinning dyon is
\[
F^{\mu\nu}_\textrm{spinning-dyon} (x) = \Re \left[ i e_2 \int \! \dd^4 q \, \del (q\cdot u_2) \,e^{iq\cdot(x-b_2)} e^{-(q \cdot a + i \theta)} \frac{ q^\mu u_2^\nu - q^\nu u_2^\mu + i \epsilon^{\mu\nu} (q, u_2)}{q^2} \right] \,.
\]
in terms of the notation of equation~\eqref{eq:epsContractionDef}; in particular,
\[
\epsilon^\mu(a,b,c) = \ep^{\mu\nu\rho\sigma} a_{\mu}b_{\nu}c_{\sigma} \,, \qquad
\epsilon^{\mu\nu}(a, b) &\equiv \epsilon^{\mu\nu\rho\sigma} a_\rho b_\sigma \,.
\]

In the electromagnetic field of the spinning dyon, the motion of our probe particle is determined by the Lorentz force
\[
\frac{\sd p^\mu}{\sd \tau} = e_1 \, F^{\mu\nu}_\textrm{spinning-dyon} (x_1(\tau)) \, u_{1 \nu}(\tau) \,.
\]
The impulse is the total time integral of this force. At leading order, the trajectory is given by equation~\eqref{eq:LOtrajectory} and we may take the
velocity $u_1$ to be constant, so we find
\[
\label{EMimpulse2}
\Delta p_1^\mu &= e_1 \int d\tau\, F^{\mu\nu}_\textrm{spinning-dyon} (b_1 + u_1 \tau) \, u_{1 \nu} \\
&= \Re \left[ i e_1 e_2 \int \! \dd^4 q \, \del(q \cdot u_1) \del(q \cdot u_2) e^{i q \cdot b} e^{-(q \cdot a + i \theta)} \frac{q^\mu (u_1 \cdot u_2) - i \epsilon^\mu(q, u_1, u_2) }{q^2} \right] \,,
\]
where $b = b_1 - b_2$ is the impact parameter.

This result for the impulse \eqref{EMimpulse2} reduces to that of \rootKerr~ \cite{Arkani-Hamed:2019ymq} when the angle 
$\theta$, and therefore the magnetic charge, is set to zero. It also reduces to the impulse due to a static dyon \cite{Huang:2019cja} when the 
spin $a$ is set to zero.

The result \eqref{EMimpulse2} was derived in the leading perturbative order (corresponding to the leading PM order in the gravitational case) but for any value of the spin parameter $a$.
It is interesting to consider the approximation where $a\ll r$, which allows an easy comparison to familiar formulae from undergraduate
electromagnetism.
In this limit, the electric and magnetic fields of a spinning dyon are given by
\begin{equation}
\begin{aligned}
F_2^{0i} &= E^i = \frac{Q}{4 \pi r^3} r^i + E^i_{\text{dipole}}
,\\
F_2^{ij} &= \ep^{ijk} B_k = \ep^{ijk}  \Big( \frac{\Qt}{4 \pi r^3} r_k + B_{k,\text{dipole}}       \Big)\,,
\end{aligned}
\end{equation}
where the electric and magnetic charges are $Q = e_2 \cos \theta$ and $\tilde Q = e_2 \sin \theta$.
The Lorentz force acting on the probe particle can then be written as
\begin{equation}\label{LorentzEM}
\frac{\sd \bp_1}{\sd\tau} =
e_1  \gamma \Big(
\bE + \bv\times\bB
\Big).
\end{equation}
Here the magnetic dipole field
\begin{equation}\label{magneticDfield}
\bB_{\text{dipole}} =  \frac{3 (\boldsymbol{\mu}_\textrm{mag} \cdot \hat{\br})  \hat{\br}}{4 \pi r^3}  -   \frac{\boldsymbol{\mu}_\textrm{mag}}{4 \pi r^3} 
\end{equation}
is given in terms of the magnetic dipole moment
\begin{equation}\label{magneticMoment}
\boldsymbol{\mu}_\textrm{mag} = e_2 \cos \theta \, \ba.
\end{equation}
Notice that this dipole moment has its origin in the spatial Newman-Janis shift!

Similarly, the electric dipole field
\begin{equation}\label{electricDfield}
\bE_{\text{dipole}} =  \frac{3 (\bae \cdot \hat{\br}) \hat{\br} }{4 \pi r^3}  -   \frac{ \bae}{4 \pi r^3} 
\end{equation}
is given in terms of the electric dipole moment
\begin{equation}
\bae = -e_2 \sin \theta \, \ba.
\end{equation}
The field of a spinning dyon, expanded in spin $a$, therefore contains four moments: electric monopole, magnetic monopoles, 
electric dipole and magnetic dipole\footnote{To all orders in spin, the field of course contains an infinite set of higher multipole moments.}.
Integrating the Lorentz force \eqref{LorentzEM} will now reproduce \eqref{EMimpulse2} with $q\cdot a \ll 1$.

\subsection{Kerr-Taub-NUT}
\label{sec:classicalKTNimpulse}

We now turn to calculate the impulse of a probe particle moving in the Kerr-Taub-NUT background. We will recover the impulse below
using the double copy at the level of scattering amplitudes, justifying at linear order the classical double copy connection between
the spinning dyon and Kerr-Taub-NUT.

Our setup is similar to the electromagnetic case.
We again take particle 1 to be a light scalar probe of mass $M_1$, and will take the Kerr-Taub-NUT itself to be heavy 
and located  at the spatial origin. The parameters of the metric are $m = G M_2 \cos \theta$ and $\ell = G M_2 \sin \theta$, where $\theta$ is 
a duality rotation angle.

It is difficult to compute the impulse on geodesics in Kerr-Taub-NUT using the metric. It is far simpler to perform the calculation by taking
advantage of the duality relation~\eqref{eq:KTNcurvature} between Kerr and Kerr-Taub-NUT, which will allow us to reduce to a computation
in a Kerr background. Since we have discussed duality at the level of curvature, we compute the impulse from the geodesic 
deviation equation. That is, we consider a family of geodesics, parameterised by a number $\lambda$, and compute the leading order impulse 
on each geodesic as a function of $\lambda$. At
zeroth order, the geodesic are all straight lines, which we take to be
\[
x(\tau, \lambda) = \lambda b + u_1 \tau \,.
\]
Thus, for $\lambda \rightarrow \infty$, the geodesics are infinitely far from the Kerr-Taub-NUT source. 
The separation vector $s^\mu$ between these geodesics (that is, the tangent vector in the $\lambda$ direction) obeys
\[
\frac{D^2 }{D\tau^2} s_\mu = R^\textrm{KTN}_{\mu\nu\rho\sigma} (x) \,u_1^\nu u_1^\rho s^\sigma\,,
\]
where $D/D\tau$ is the covariant differential in the $\tau$ direction. Since we work at linearised order, we may replace $D/D\tau$ with $\sd/\sd\tau$ on
left-hand side of this equation; on the right-hand side, we may take the velocity vectors to be constant, and evaluate the curvature on the 
unperturbed straight line trajectories. We may also take $s^\mu = \sd x^\mu(\tau, \lambda) / \sd\lambda$. Thus the deviation equation is
\[
\frac{\sd^3}{\sd \lambda\, \sd\tau^2} x_\mu = R^\textrm{KTN}_{\mu\nu\rho\sigma}(\lambda b + u_1\tau) \, u_1^\nu u_1^\rho  b^\sigma\,.
\]
We will focus specifically on the impulse on the $\lambda = 1$ geodesic. It is helpful to note that 
since the curvature tensors fall off as $r^{-3}$, there is no impulse on the $\lambda \rightarrow \infty$ geodesic. As a result we can
compute the impulse on our $\lambda = 1$ geodesic by integrating the deviation equation twice:
\[
\label{eq:lambdaRegion}
M_1 \int_{-\infty}^\infty \sd\tau \int_\infty^1 \sd \lambda \frac{\sd^3}{\sd \lambda\, \sd\tau^2} x_\mu &= M_1 \int_{-\infty}^\infty \sd\tau \left. \frac{\sd^2 x_\mu}{\sd\tau^2} \right|_{\lambda = 1} - M_1 \int_{-\infty}^\infty \sd\tau \left. \frac{\sd^2 x_\mu}{\sd\tau^2} \right|_{\lambda \rightarrow \infty} \\
&= M_1 \Delta u_{1\mu} \Big|_{\lambda = 1} \\
&= \Delta p_{1\mu} \,.
\]
In other words, the impulse can be obtained by integrating over the curvature
\[
\Delta p_{1\mu} 
&= M_1 \int_\infty^1 \sd \lambda \int_{-\infty}^\infty \sd\tau \, R^{\textrm{KTN}}_{\mu\nu\rho\sigma}(\lambda b + u_1\tau)\,\, u_1^\nu u_1^\rho  b^\sigma\,.
\]

Now we can make good use of the duality relation~\eqref{eq:KTNcurvature} between the linearised curvatures of Kerr-Taub-NUT and Kerr. 
Defining
\[
{\Theta}^+_{\mu\nu\rho\sigma} = \Theta^\textrm{Kerr}_{\mu\nu\rho\sigma} + i \, \tilde \Theta^\textrm{Kerr}_{\mu\nu\rho\sigma} \,,
\]
the impulse becomes
\[
\label{eq:impulseFromDeviation}
\Delta p^\mu_{1} = GM_1M_2 \Re e^{-i \theta} \int_\infty^1 \sd \lambda \int_{-\infty}^\infty \sd\tau \, \Theta^{+\mu}{}_{\nu\rho\sigma}(\lambda b + u_1\tau)\, \, u_1^\nu u_1^\rho  b^\sigma\,.
\]
This relation achieves our goal of reducing the problem of finding the impulse in Kerr-Taub-NUT to an integral over the simpler curvature of Kerr.

To evaluate equation~\eqref{eq:impulseFromDeviation}, it is most convenient to work in the tetrad formalism and express curvature in terms of spin connections. At the linearized level, one can formulate the Riemann tensor in terms of the spin connection as follows\footnote{We do not distinguish between indices in the frame bundle and on the manifold in this case as they are the same at the linear level.}
\begin{equation}
\frac1{m} R^\textrm{Kerr}_{\mu\nu\rho\sigma} = \Theta^\textrm{Kerr}_{\mu\nu\rho\sigma} = \pd_\rho\omega_{\mu\nu\sigma} - \pd_\sigma\omega_{\mu\nu\rho},
\end{equation}
where the spin connection itself is\footnote{We define $x_{[\alpha} y_{\beta]} \equiv \frac12 (x_\alpha y_\beta - x_\beta y_\alpha)$. More details can be found in appendix~\ref{sec:conventions}.}
\begin{equation}
m \, \omega_{\mu\nu\rho} = \pd_{[\nu}h_{\mu]\rho} \,.
\end{equation}
We have factored the linear dependence on the mass from the spin connection for convenience.
As the Riemann tensor is linear in the spin connection, the dual tensor is then simply given by dualizing the latter:
\[
\tilde{\Theta}^\textrm{Kerr}_{\mu\nu\rho\sigma} &= \pd_\rho\tilde{\omega}_{\mu\nu\sigma} - \pd_\sigma\tilde{\omega}_{\mu\nu\rho} \,,\\
\]
where
\[
\tilde\omega_{\mu\nu\rho} &= \frac12 \epsilon_{\mu\nu}{}^{\alpha\beta} \omega_{\alpha\beta\rho} \,.
\]

In terms of the spin connection, we can immediately perform the $\lambda$ integration in the impulse~\eqref{eq:impulseFromDeviation} 
as follows. Let us write the self-dual projection of the spin connection as
\[
\omega^+_{\mu \nu \rho} = \omega_{\mu\nu\rho} + i \tilde \omega_{\mu\nu\rho} \,,
\]
so that
\[
\Delta p_{1\mu}&= GM_1 M_2 \Re e^{-i \theta} \int_\infty^1 \sd \lambda \int_{-\infty}^\infty \sd\tau \, \left(\pd_\rho \omega^+_{\mu\nu\sigma} (\lambda b + u_1\tau)-\pd_\sigma \omega^+_{\mu\nu\rho} (\lambda b + u_1\tau)\right) u_1^\nu u_1^\rho b^\sigma \\
&= GM_1M_2 \Re e^{-i \theta} \int_\infty^1 \sd \lambda \int_{-\infty}^\infty \sd\tau \, \left(\frac{d}{d\tau} \omega^+_{\mu\nu\sigma} (\lambda b + u_1\tau) u^\nu b^\sigma -\frac{d}{d\lambda} \omega^+_{\mu\nu\rho} (\lambda b + u_1\tau)u_1^\nu u_1^\rho\right)  \,. \\
\]
This expression simplifies considerably. The spin connection falls off as $r^{-2}$, so we may drop the total $\tau$ derivative: it evaluates to the 
spin connection evaluated at infinity. Similarly, the total $\lambda$ derivative evaluates to the $\omega^+_{\mu\nu\rho} (b + u_1 \tau) - 0$. The 
$-0$ here computes the vanishing deviation of the $\lambda \rightarrow \infty$ geodesic, consistent with equation~\eqref{eq:lambdaRegion}.

We have arrived at
\[
\Delta p_{1\mu} &= -GM_1M_2 \Re e^{-i \theta} \int_{-\infty}^\infty \sd\tau \,  \omega^+_{\mu\nu\rho} (b + u_1\tau) \, u_1^\nu u_1^\rho \\
&= -GM_1M_2 \Re e^{-i \theta} \int_{-\infty}^\infty \sd\tau \left( \omega_{\mu\nu\rho} (b + u_1\tau) + i \tilde\omega_{\mu\nu\rho} (b + u_1\tau) \right) \, u_1^\nu u_1^\rho \,,
\]
so the impulse is simply given by integrating the geodesic equation with a spin connection given by duality. This would also be a reasonable
starting point for a computation of the impulse, but we felt it was worth making the link to the Riemann tensor explicitly.

Our goal now is to evaluate this impulse. We will break the computation up into two parts: the ``electric'' impulse arising from 
$\omega_{\mu\nu\rho}$ (which is the impulse on a probe in Kerr) and the ``magnetic'' impulse due to the dual spin connection. That is,
\[
\label{eq:totalImpulse}
\Delta p_{1\mu} = -m M_1 \int_{-\infty}^\infty \sd\tau\,  \omega_{\mu\nu\rho} (b + u_1\tau) \, u_1^\nu u_1^\rho - \ell M_1 \int_{-\infty}^\infty \sd\tau \, \tilde \omega_{\mu\nu\rho} (b + u_1\tau) \, u_1^\nu u_1^\rho  \,. 
\]
Since the on-shell amplitude approach that we will study in the next section yields a momentum space amplitude, we find it convenient to work directly in momentum space and perform the dualization there. It is helpful to exploit the kinematics of the problem to simplify the final result
(and to facilitate a comparison to scattering amplitudes). We discuss the details of the kinematics in appendix~\ref{sec:kinematics}; here, we
will be content to use the various results from the appendix when we need them. 

For our purposes, it is useful to work with the linearized Kerr graviton (for finite Kerr parameter $a$) in de Donder gauge. A useful
expression was given by Vines in reference \cite{Vines:2017hyw}:
\begin{align}
h_{\mu\nu}^\textrm{Kerr} &= \cl{P}_{\mu\nu\alpha\beta}\left(u_2^\alpha u_2^\beta \cos(a\cdot\pd) + u_2^{(\alpha}\epsilon^{\beta)}{}_{\rho\sigma\lambda}u_2^\rho a^\sigma\pd^\lambda\frac{\sin(a\cdot\pd)}{a\cdot\pd}\right)\frac{4m}{r},
\end{align}
where $u_{2\mu}$ is defined via $u_2\cdot\pd = \pd_t$, and the trace reverser is defined as
\begin{equation}
\cl{P}_{\mu\nu\alpha\beta} \coloneqq \frac12\left(\eta_{\mu\alpha}\eta_{\mu\beta} + \eta_{\mu\beta}\eta_{\mu\alpha} - \eta_{\mu\nu}\eta_{\alpha\beta}\right).
\end{equation}

As in the electromagnetic case, it is useful to Fourier transform to momentum space. Denoting the momentum-space form of the spin connection
as $\Omega$, we have
\[
\label{KerrSpinConn}
m \,& \Omega_{\mu\nu\rho}(q) = -i\int \sd^4x \, e^{-i{q}\cdot{x}}q_{[\mu}h^\textrm{Kerr}_{\nu]\rho}(x) \\[0.4em]
&= -i\int \sd^4x~e^{-i{q}\cdot{x}}q_{[\mu}P_{\nu]\rho\alpha\beta}\left(u_2^\alpha u_2^\beta \cosh(a\cdot q) +i u_2^{(\alpha}\epsilon^{\beta)}(u,a,q)\frac{\sinh(a\cdot q)}{a\cdot q}\right)\frac{4m}{r} \\[0.4em]
&= -i\del(q \cdot u_2)\left((2q_{[\mu}u_{2\nu]}u_{2\rho} + q_{[\mu}\eta_{\nu]\rho}) \cosh(a\cdot q) + 2iq_{[\mu}u_{2(\nu]}\epsilon_{\rho)}(u_2,a,q)\frac{\sinh(a\cdot q)}{a\cdot q}\right)\frac{8\pi m}{q^2}.
\]
Writing the electric part of the impulse as
\begin{equation}
\Delta p_{1, \text{elec}}^\mu = -m M_1 \int_{-\infty}^\infty \sd\tau\,  \omega^\mu{}_{\nu\rho} (b + u_1\tau) \, u_1^\nu u_1^\rho ,
\end{equation}
we see that it can be determined from the momentum space spin connection as follows
\begin{equation}\label{momimpulse}
\Delta p_{1, \text{elec}}^\mu = -mM_1\int \sd\tau \int \dd^4q\,e^{iq\cdot (b+u_1 \tau)}~\Omega_{~\nu\rho}^{\mu}(q) \, u_1^\nu u_1^\rho \;.
\end{equation}
Plugging eq. \eqref{KerrSpinConn} into eq. \eqref{momimpulse}, we find that the electric impulse is
\[
\Delta p_{1, \text{elec}}^\mu 
&=
-mM_1\int \sd\tau \int \dd^4q\,e^{i{q}\cdot {(b+u_1 \tau)}}~\Omega_{~\nu\rho}^{\mu}(q)\, u_1^\nu u_1^\rho
\\
&=  4\pi i \, m M_1 \int \dd^4q~\hat{\delta}(q\cdot u_1)\hat{\delta}(q\cdot u_2)e^{iq\cdot b}\frac{q^\mu}{q^2}
\\&
\qquad \times\bigg[(2(u_1\cdot u_2)^2 - 1)\cosh(q\cdot a) + 2i(u_1\cdot u_2)\frac{\epsilon(u_1,u_2,a,q)}{q\cdot a}\sinh(q\cdot a)\bigg]
\\ &=  4\pi i \, m M_1 \int \dd^4q~\hat{\delta}(q\cdot u_1)\hat{\delta}(q\cdot u_2)\frac{e^{iq\cdot b}}{q^2}
\\&\qquad\times
\bigg[q^\mu\cosh 2w\cosh(q\cdot a) - 2i\epsilon^\mu(u_1,u_2,q)\cosh w\sinh(q\cdot a) + \mathcal{O}(q^2)\bigg].
\label{eq:electricImpulse}
\]
In arriving at this result, we exploited the kinematical discussion in appendix~\ref{sec:kinematics} to make the
replacement~\eqref{eq:qReplacement}. We have also ignored terms proportional to $q^2$ in the
square brackets of equation~\eqref{eq:electricImpulse}. These terms cannot contribute to the impulse, as we explain
in detail in appendix~\ref{sec:integrals}. We will systematically neglect any terms of this form below.

The total impulse imparted to a probe particle in the Kerr-Taub-NUT background, is the sum~\eqref{eq:totalImpulse} of this electric impulse and a magnetic contribution due to the dualized piece of the spin connection.
Transforming to momentum space, the dualized spin connection is given by 
\[
\label{eq:dualSpinConnectionStep}
\tilde{\Omega}_{\mu\nu\rho}(q) &= \frac{1}{2}\epsilon_{\mu\nu}^{~~\alpha\beta}\Omega_{\alpha\beta\rho}(q) \nn\\
&= -\frac{4\pi i}{q^2}\hat{\delta}(q\cdot u_2)\bigg[\left(2\epsilon_{\mu\nu}(q,u_2)u_{2\rho} - \epsilon_{\mu\nu\rho}(q)\right)\cosh(q\cdot a)  \nn\\
&~~~~~~~~~~+ i\left(\epsilon_{\mu\nu}(q,u_2)\epsilon_\rho(u_2,a,q) + 2(q\cdot a)q_{[\mu}u_{2\nu]}u_{2\rho}\right)\frac{\sinh(q\cdot a)}{q\cdot a}\bigg].
\]
We simplified this expression with the help of the result
\begin{equation}
\epsilon_{\mu\nu}{}^{\beta}(q)\epsilon_{\beta}(u,a,q) = 2u_{[\mu}q_{\nu]}(q\cdot a) + 2a_{[\mu} u_{2\nu]} q^2 \,,
\end{equation}
which follows by expanding the contraction of two Levi-Civita tensors, and setting $q \cdot u = 0$ in view of the delta function in 
equation~\eqref{eq:dualSpinConnectionStep}. As a result, the magnetic contribution to the impulse~\eqref{eq:totalImpulse} is
\[
\Delta p^\mu_{1\textrm{, mag}} &= - \ell M_1 \int_{-\infty}^\infty \sd\tau \, \tilde \omega^\mu{}_{\nu\rho} (b + u_1\tau) \, u_1^\nu u_1^\rho \\
&= 4\pi i \, \ell M_1  \int \dd^4q~\hat{\delta}(q\cdot u_1)\hat{\delta}(q\cdot u_2)e^{iq\cdot b}\frac{1}{q^2} \left[ 2\epsilon^\mu(u_1, u_2, q) \cosh w \cosh q\cdot a  \phantom{ \frac11}\right. \\
& \hspace{0.17\textwidth}
\left. 
+ i \left(-\epsilon^\mu(u_1, u_2, q) \frac{\epsilon(u_1, u_2, a, q)}{q\cdot a} + q^\mu \cosh^2w \right) \sinh q\cdot a \right] \,.
\]
A considerable simplification now occurs if we once again use equation~\eqref{eq:qReplacement} to replace the vector 
$\epsilon^\mu(u_1, u_2, q)$ with $q^\mu$ in the last line. In doing so, we encounter the Gram determinant $\epsilon(a, q, u_1, u_2)^2$
which is evaluated in appendix~\ref{sec:kinematics}. We again ignore powers of $q^2$, and find
\[
\label{eq:mangeticImpulse}
\Delta p^\mu_{1\textrm{, mag}} =
4\pi i \, \ell M_1  \int \dd^4q~\hat{\delta}(q\cdot u_1)\hat{\delta}(q\cdot u_2)e^{iq\cdot b}\frac1{q^2} &\left[ 2 \epsilon^\mu(u_1, u_2, q) \cosh w \cosh q\cdot a \right. \\
&\hspace{40pt}\left. + i q^\mu \, \cosh 2w \sinh q \cdot a \right] \,.
\]

The total Kerr-Taub-NUT impulse is the sum of the electric~\eqref{eq:electricImpulse} and magnetic~\eqref{eq:mangeticImpulse} contributions, namely
\begin{equation}
\begin{aligned}
\Delta p^\mu_1 
&= 4\pi i\, M_1\int \dd^4q~\hat{\delta}(q\cdot u_1)\hat{\delta}(q\cdot u_2)e^{iq\cdot b}\frac1{q^2}
\\&
\qquad \times
\Big[ \left( m\,  q^\mu \cosh2w +2\ell\, \epsilon^\mu(u_1,u_2,q) \cosh w\right)\cosh q\cdot a
\\
&\qquad \qquad   + \left(  2i m\, \epsilon^\mu(u_1,u_2,q) \cosh w  +i \ell \, q^\mu \cosh 2w\right)\sinh(q\cdot a)\Big] .
\end{aligned}
\end{equation}
For later comparison to scattering amplitudes, it will be useful to rewrite this result in a slightly different form. We recall that the mass
and NUT parameters are
\[
m = GM_2 \cos \theta\, , \quad \ell = GM_2 \sin \theta \,,
\]
in terms of the duality angle $\theta$. Bearing this in mind, simple trigonometry allows us to write the impulse as
\[
\label{gravityImpulse}
\Delta p_1^\mu = \Re 4\pi GM_1 M_2 \int \! \dd^4 q & \, \del(q \cdot u_1) \del(q \cdot u_2) \frac{e^{iq \cdot b}}{q^2} \\
&\times \left[ iq^\mu \, \cosh 2w -2 \epsilon^\mu(u_1, u_2, q) \cosh w \right]  e^{-(q\cdot a + i \theta)} \,.
\]
It is startling how simple the final result is, particularly in view of the complicated form of the linearised Kerr-Taub-NUT 
metric, equation~\eqref{eq:linKTNmetric}. We will soon see that scattering amplitudes make this simple form almost obvious
from the beginning.

\subsection{Geodesics}

In the previous subsection we have computed the impulse in the Kerr-Taub-NUT background directly, by dualizing the Kerr impulse. In this subsection we will study the geodesic equation in non-relativistic form to bring out the physics more clearly. We will also expand in spin $a$
in the subsection, again to make the physics more transparent.

A test particle of mass $M_1$ is described by its four-momentum
\begin{equation}
p^{\mu}_1= M_1 u^{\mu}_1 \,,
\end{equation}
which is given in terms of the four-velocity
\begin{equation}
u^{\mu}_1=\frac{\sd x^{\mu}_1 }{\sd\tau} = \gamma \left(1,\bv\right) \,.
\end{equation}
The motion of the massive particle is described by the geodesic equation
\begin{equation}
\frac{\sd^2 x^{\mu}_1}{\sd\tau^2} = - \Gamma ^{\mu}_{\nu \rho} \frac{\sd x^{\nu}_1}{\sd\tau}\frac{\sd x^{\rho}_1}{\sd\tau} \,,
\end{equation}
so that the spatial components of the force are
\begin{equation}\label{geoEq}
\frac{\sd  \bp _1  }{\sd\tau} = 
- M_1   \gamma ^2\Big(  \Gamma ^{i}_{00} 
+2 \Gamma ^{i}_{0j} v^j
+ \Gamma ^{i}_{j k} v^j v^k 
\Big)
\,.
\end{equation}
We have used the familiar expression $\gamma = \sd t/\sd \tau = \left( 1-\bv\cdot \bv \right)^{-1/2}$. We now wish to evaluate the geodesic equation on the Kerr-Taub-NUT background \eqref{KTNmetric} in the leading post-Minkowskian approximation.
For simplicity we will also take now the spin parameter to be very small $a\ll r$.
In this case the metric can be expanded around flat spacetime as
\begin{equation}
g_{\mu\nu} = \eta_{\mu\nu} + h_{\mu\nu},
\end{equation}
where $\eta_{\mu\nu} $ is the flat metric and the leading correction in Cartesian coordinates is given by
\begin{equation}\label{NPMmetric}
\begin{aligned}
h_{00}&= \frac{2m}{r}+ \frac{2 \Jt  \cos \theta }{r^2}=\frac{2m}{r}+ \frac{2 \Jt  z }{r^3},  \\
h_{0i} &=  \frac{2}{r^3 } \Big( \ell \frac{z}{(1 -  \frac{z^2}{r^2})}  + 2J \Big)
\left(
\begin{tabular}{c}
$+y$   \\
$-x$    \\
$0$  
\end{tabular}
\right),
\\
h_{ij} &= \frac{2mx_ix_j}{r^3}+ \frac{2 \Jt z}{r^3} \delta_{ij}.
\end{aligned}
\end{equation}
where
\begin{equation}
\begin{aligned}
J &= a m , \\
\Jt &= a \ell ,
\end{aligned}
\end{equation}
are the angular momentum and dual angular momentum aspects. 
In polar coordinates we have $h_{0i} \,\sd x^i = -2 \ell \cos \theta \,\sd \varphi - \frac{4J}{r} \sin ^2  \theta \, \sd \varphi$.
The Christoffel symbols appearing in the geodesic equation \eqref{geoEq} then take the form
\begin{equation}
\begin{aligned}
\Gamma ^i_{00} &= - \frac{1}{2} \pa^i h_{00} , \\
\Gamma ^i_{0j} &= \frac{1}{2} \eta^{ik} \left(\pa_k h_{0j} - \pa_j h_{0k} \right),\\
\Gamma^i _{jk} &= m\left( \frac{2x^i}{r^3} \delta_{jk}  - \frac{3 x^i x_j x_k}{r^5} \right)\\
&+\frac{\Jt}{r^3} \delta^i_{j}\left[\delta_{k}^3-\frac{3z}{r^2}x_{k}\right]
+\frac{\Jt}{r^3} \delta^i_{k}\left[\delta_{j}^3-\frac{3z}{r^2}x_{j}\right]
-\frac{\Jt}{r^3} \delta_{jk}\left[\delta^{i3}-\frac{3z}{r^2}x^i\right]
.
\end{aligned}
\end{equation}
Notice that the first and third terms in the geodesic equation \eqref{geoEq} are proportional to the mass aspect of the background metric and the dual angular momentum while the second term is proportional to the NUT charge and the angular momentum aspect. We will soon see that accordingly the first and third terms will contribute to the ``electric'' component of the force while the second term will contribute to the ``magnetic'' component.

We can bring the geodesic equation to a form similar to the Lorentz force in gauge theory \eqref{LorentzEM} by defining
\begin{equation}
\begin{aligned}
\phi & \equiv - \frac{1}{2}h_{00}, \\
A_i &\equiv - \frac{1}{4}h_{0i}.
\end{aligned}
\end{equation}
The potentials $\phi$ and $\bA$ are the analogues of the scalar and vector potentials in electrodynamics.
Explicitly, the resulting potentials are given by
\begin{equation}
\begin{aligned}
\phi  &= - \frac{m}{r}  -\frac{ \bae \cdot \hat{\br}}{4\pi r^2}  , \\
\bA &=\bA _{\text{Dirac}}+\bA _{\text{Gauge}}+\bA _{\text{dipole}} .
\end{aligned}
\end{equation}
Here, the electric dipole moment is proportional to the dual angular momentum
\begin{equation}
\bae = - 4\pi \Jt \hat{\bz} .
\end{equation}
The vector potential receives three distinct contributions; the first is the Dirac potential, given by
\begin{equation}
\bA_{\text{Dirac}} = \frac{2\ell}{r} \frac{\br \times \bn}{r - (\br\cdot \bn)}.
\end{equation}
The second contribution, given by
\begin{equation}
\bA_{\text{Gauge}} = \frac{2 \ell}{r^2 -z^2} ( \br \times \bn ),
\end{equation}
is a pure gauge whose magnetic field $\bD \times \bA_{\text{Gauge}} $ vanishes. Finally, the magnetic dipole potential
\begin{equation}
\bA_{\text{dipole}} = \frac{\ba \times \hat{\br}}{4\pi r^3}
\end{equation}
is given in terms of the magnetic dipole moment
\begin{equation}
\ba= -16 \pi  J \hat{\bz},
\end{equation}
which is proportional to the angular momentum aspect.

We can then write the geodesic equation \eqref{geoEq} in terms of the scalar and vector potentials as follows
\begin{equation}\label{FinalGeo}
\frac{\sd \bp _1}{\sd\tau} =
M_1 \gamma \Big(
\frac{2\gamma ^2 -1}{\gamma} 
\bE
+4 \gamma \,  \bv\times\bB
+\bF_T
+ \bF_v
\Big),
\end{equation}
where the electric and magnetic fields are given by
\begin{equation}\label{fields}
\begin{aligned}
\bE& =\bD \phi= \bE_{\text{monopole}}+\bE_{\text{dipole}}  ,\\
\bB&= \bD  \times \bA= \bB_{\text{monopole}} +\bB_{\text{dipole}}.
\end{aligned} 
\end{equation}
The different field components are given by the same precise form as their electromagnetic analogues; the monopole fields are
\begin{equation}
\begin{aligned}
\bE_{\text{monopole}} &=  \frac{m}{r^3} \br  \\
\bB_{\text{monopole}}   &=\frac{2\ell}{r^3} \br,
\end{aligned}
\end{equation}
and the dipole fields are
\begin{equation}\label{MFangular}
\begin{aligned}
\bE_{\text{dipole}} &=  \frac{3 (\bae \cdot \hat{\br}) \hat{\br} }{4 \pi r^3}  -   \frac{ \bae}{4 \pi r^3} ,\\
\bB_{\text{dipole}} &= \frac{3 (\ba \cdot \hat{\br}) \hat{\br}}{4\pi r^3} - \frac{\ba}{4 \pi r^3}.
\end{aligned}
\end{equation}

In addition to the electric and magnetic forces we also find transient and velocity forces. The transient force is given by
\begin{equation}
\left( \bF_T \right)^i = \frac{\sd}{\sd \tau} \Big( \frac{1}{2 \gamma} \eta^{ij} v^k h_{jk} \Big),
\end{equation}
which can be expressed in terms of the electric field
\begin{equation}
\bF_T= \frac{\sd}{\sd\tau}\Big( \frac{  \left(\bv \cdot \br\right)  }{\gamma ^2} \bE \Big).
\end{equation}
The velocity force is given by
\begin{equation}
\bF_v =  - \frac{1}{\gamma} \frac{m (  \bv \cdot \br )  }{  r^3 }  \bv \, ,
\end{equation}
and is proportional to the mass aspect. In the above derivation we have used the fact that the velocity is constant at leading order in the post-Minkowskian approximation.

The geodesic equation \eqref{FinalGeo} was studied in detail in \cite{Huang:2019cja}. The new contributions here are due to the electric and the magnetic dipole moments, but the form of the geodesic equation in terms of the fields $\bE,\bB,\bF_T$ and $\bF_v$ is the same as in \cite{Huang:2019cja}. Let us summarize some of the main features of the resulting force \eqref{FinalGeo}.
First of all, in the post-Newtonian (PN) approximation, where velocities are small ($\gamma \rightarrow 1$), it reduces to the form of the Lorentz force in electrodynamics \eqref{LorentzEM}
\begin{equation}
\frac{\sd \bp _1}{\sd\tau} \Big| _{\text{PN}}=
M_1 \Big(
\bE +4   \bv\times\bB
\Big),
\end{equation}
where the charge of the test particle is replaced by minus its mass.
At this point we would like to emphasize that nowhere in this paper we consider the velocity of the probe particle to be small.
We use this limit as a check that our formula \eqref{FinalGeo} reduces to the known form of the force in the PN approximation, including the factor of 4 in front of the magnetic force (see, for example, \cite{DeWitt:1966yi,Wald:1984rg}).
In particular, note that both the transient and velocity forces are negligible to leading order in the PN approximation.
More generally, when the velocity of probe particles is not assumed to be small, the electric and magnetic forces appear with coefficients that depend on the Lorentz factor $\gamma$ and there are two additional forces. The transient force $\bF_T$ is a total derivative with respect to $\tau$ and therefore its contribution to the impulse vanishes since the force decays to zero at the boundaries. The velocity force $\bF_v$ is not a total derivative but its contribution to the impulse is zero as well (see \cite{Huang:2019cja}).
Both forces $\bF_T$ and $\bF_v$ represent transient, short-lived, effects that do not contribute to the impulse.

Finally, we express the $\gamma$-dependent coefficients in \eqref{FinalGeo} in terms of the rapidity $w$, defined by
\begin{equation}
\begin{aligned}
\gamma &= \cosh w , \qquad   &  \sqrt{\gamma^2-1} = \sinh w, \\
2\gamma ^2 -1  &= \cosh 2w, \qquad   &  2\gamma \sqrt{\gamma^2-1} = \sinh 2w ,
\end{aligned}
\end{equation}
such that the geodesic equation now takes the form
\begin{equation}\label{FinalGeow}
\frac{\sd \bp _1 }{\sd\tau} =
M_1 \gamma \Big(
\frac{\cosh 2w }{\cosh w}  
\bE
+2 \frac{\sinh 2w }{\sinh w}   \bv\times\bB
+\bF_T
+ \bF_v
\Big).
\end{equation}
The double-copy structure relates gauge theory amplitudes with rapidity $w$ to those in gravity with a doubled rapidity $2w$.
As in \cite{Huang:2019cja}, we see that the double-copy structure is realized in the equations of motion via the coefficients $\frac{\cosh 2w }{\cosh w}$ and $\frac{\sinh 2w }{\sinh w}$.

\section{A Network of Amplitudes}\label{sec:amplitudes}

So far, we have justified the ``double copy'' axis of figure~\ref{fig:solutionCube} by appealing to previous work~\cite{Luna:2018dpt} on the
application of the double copy to classical solutions. In this section, we take a significant new step by finding amplitudes associated
with each of our solutions. As we will see, these amplitudes double-copy precisely when we claim the solutions do so.

\subsection{Connecting Amplitudes to Solutions}

We begin by assigning a three-point amplitude to each node in the network of solutions, making use of knowledge the community has
gathered about how duality~\cite{Caron-Huot:2018ape,Huang:2019cja,Moynihan:2020gxj}, the
Newman-Janis shift~\cite{Arkani-Hamed:2019ymq}, and the double copy~\cite{Kawai:1985xq,Bern:2008qj,Bern:2010ue,Bern:2010yg} act on amplitudes. Our starting point is the simplest solution: the Coulomb charge.

We will use the massive four-dimensional spinor formalism developed in Ref. \cite{Arkani-Hamed:2017jhn}. This language allows us to discuss
any massive amplitude in four dimensions, and will be particularly helpful below to construct the four point amplitudes which describe
scattering at leading order.

Three-particle amplitudes are constructed from $SL(2,\mathbb{C})$ (Lorentz) invariants, but nevertheless they have little group charge (except
for the case of a purely scalar interation). We use spinors to carry the little group weight. For many three point amplitudes, a simple basis
of spinors exists~\cite{Arkani-Hamed:2017jhn}. For examples (like our Coulomb amplitude) involving a massless line, we may write the massless momentum as $\lambda_\alpha \tilde \lambda_{\dot \alpha}$. 
Then we build our basis from $\lambda^\alpha$ and $\lambda_{\dot \alpha} p_1^{\dot \alpha \alpha}$ where $p_1$ is the momentum of one of
the massive legs at the vertex. 

In the case at hand, however, the two massive lines in the amplitude have the same mass,
which leads to a degeneration in this basis. In fact,
a basis of linearly independent massless spinors cannot be constructed. Instead, we can define their constant of proportionality $x$~\cite{Arkani-Hamed:2017jhn}
\begin{equation}\label{xDef}
x_1\lambda^\alpha \ = \ \frac{\tilde{\lambda}_{\dot{\alpha}}p_1^{\dot{\alpha}\alpha}}{M_1} \:,\qquad \frac{1}{x_1}\tilde{\lambda}^{\dot{\alpha}} \ = \ \frac{p_1^{\dot{\alpha}\alpha}\lambda_\alpha}{M_1} \;,
\end{equation}
The so-called $x$-factor carries the little group weighting of the propagating massless particle. 

The Coulomb amplitude is simply this $x$ factor up to normalisation:
\[
\label{eq:CoulombAmplitude}
\mathcal{A}_3^\pm 
&= \sqrt{2} e M x^\pm \,,
\]
where the charge is $e$. Notice that the power of the $x$ factor controls the helicity of the photon at the vertex.

With this starting point, we can use our three operations to take us on a journey visiting each node of figure~\ref{fig:solutionCube}, 
assigning an amplitude to each node. We just need to understand how the operations act on amplitudes. The double copy is easy:
the only available dynamical object is $x$, and in the gravitational amplitude the factor must appear with power $\pm 2$ depending
on the graviton polarisation. Thus, 
up to a constant of proportionality, the gravitational amplitude $\mathcal{M}_3$  
is simply the square of  $\mathcal{A}_3$:
\[
\mathcal{M}_3 = \frac{\kappa}{2} M^2 x^{\pm 2} \,.
\]

The action of duality on a gauge amplitude is also straightforward~\cite{Huang:2019cja,Moynihan:2020gxj}. 
As we have seen in earlier sections of this paper, duality rotates the self-dual (or anti-self-dual) curvature by a phase.
Three-point helicity amplitudes are themselves chiral objects, so they also pick up a phase under duality:
\[
\label{eq:dualityPhase}
\mathcal{A}_3^{\pm}{}' = e^{\pm i \theta} \mathcal{A}_3^{\pm} \,.
\]
This holds for both electromagnetic and gravitational amplitudes. Our application of this operation will be to purely classical 
observables, but we note that reference~\cite{Csaki:2020inw} introduced a new approach to magnetically charged amplitudes.

It remains to see how the Newman-Janis shift acts on amplitudes. The shift was recently understood beginning with the electromagnetic 
amplitude for a massive spin $S$ particle using the formalism of massive spinor-helicity in four dimensions~\cite{Arkani-Hamed:2017jhn}. 
Quite remarkably, in the limit where $\hbar$ is taken to zero while $S$ the spin angular momentum $S \hbar$ is held fixed, the
amplitude becomes
\[
\label{eq:NJphase}
\mathcal{A}_3^{\pm}{}' = e^{\pm q \cdot a} \mathcal{A}_3^{\pm} \,,
\]
where $a$ is a vector parameterising the spin (see \cite{Aoude:2020onz} for a finite spin approach). We are unaware of a simple a 
priori argument for this action. The NJ shift in gravity
is again simply the double copy of the shift in EM. 

\begin{figure}[t]
	\begin{center}
		\includegraphics[width=1\textwidth]{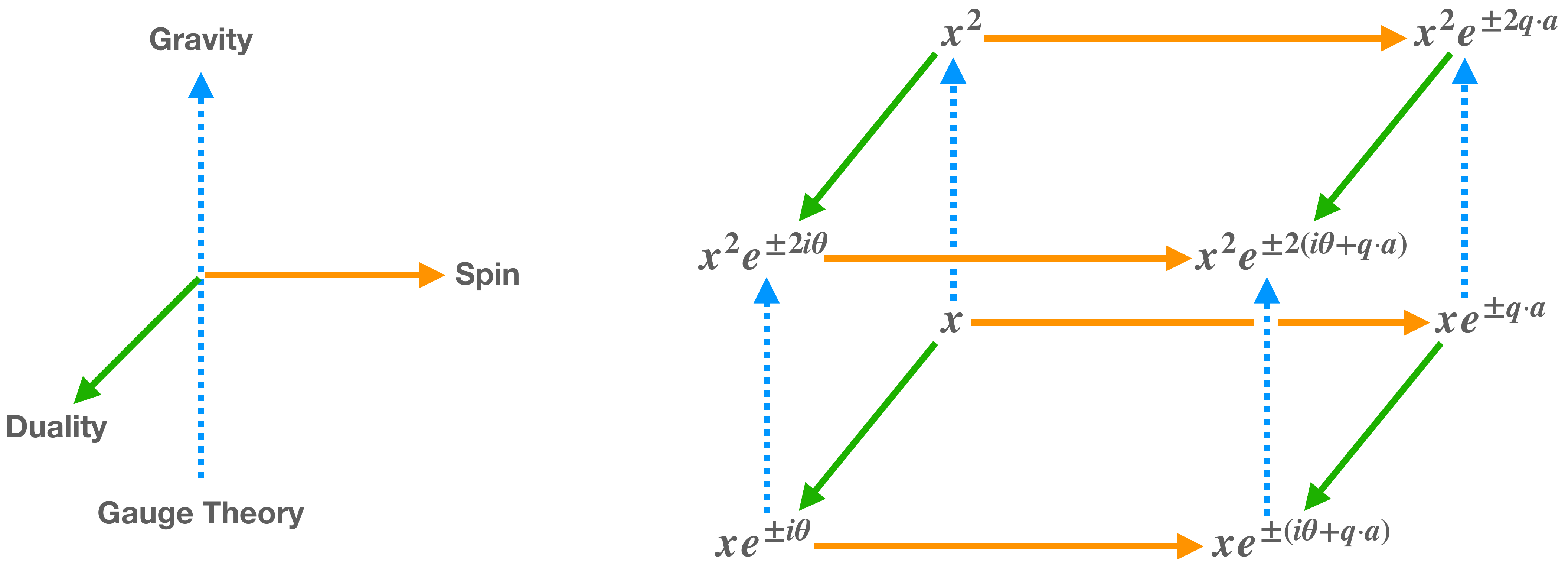}
	\end{center}
	\caption{Three-particle amplitudes related by the double copy (dashed blue arrow), duality (green arrow) and the Newman-Janis shift (orange arrow).
For brevity, throughout the paper we have rescaled the gravitational phase factors by a factor of half.
}
	\label{fig:amplitudeCube}
\end{figure}

Starting from the Coulomb case, it is now an easy matter to write down amplitudes associated with each node of our network. The
result is figure~\ref{fig:amplitudeCube}. In the figure, we strip the coupling constants from the Coulomb amplitude for clarity, leaving the 
core dynamics in
the factor $x$. Green arrows in the figure are associated with duality rotation, so the introduce factors of $e^{\pm i \theta}$ (or its square
in gravity). Orange arrows introduce the spin factor $e^{\pm q \cdot a}$ (or the square). Meanwhile the dotted blue arrows implement the 
double copy by squaring the amplitudes. We also have written the gravitational duality and Newman-Janis phases with a factor 2 so that
the double copy is manifest. This choice is purely a convention; for notational simplicity we will omit the factor 2 in the explicit computations
below.

So we have a unique assignment of amplitude to each of our solutions. The amplitudes, by construction, double-copy from gauge
theory to gravity as they should. We can now use these amplitudes to check that the solutions which we claim are related by
the double copy indeed behave as they should. Specifically, we will recompute the impulses of section~\ref{sec:impulse} using the
methods of scattering amplitudes. We will obtain a precise match between impulses between pairs of solutions related by
the classical double copy, and impulses computed using the double copy of amplitudes.

\subsection{Electromagnetic Impulses from Amplitudes}

We begin with a discussion of impulse and amplitudes in the electromagnetic case. This will be a useful warm-up for the (similar)
gravitational case, and also provides a sanity check that duality and the Newman-Janis shift act as expected on our three-point
amplitudes.

In the electromagnetic case, we scatter a probe charge off our electromagnetic backgrounds. One can readily determine the leading
order impulse imparted on the scalar probe particle via~\cite{Kosower:2018adc}
\begin{equation}\label{impulseamp}
\Delta p^\mu_1 = \frac{1}{4M_1M_2}\int \dd^4q \,\hat{\delta}(q\cdot u_1)\hat{\delta}(q\cdot u_2)e^{iq\cdot b}\, iq^\mu \cl{A}_4(1,2\rightarrow 1',2')\big\vert_{q^2\rightarrow 0} \;,
\end{equation}
where $u_1$ and $u_2$ are the proper velocities of particles 1 (the probe) and 2 (the classical source), and $\cl{A}_4$ is the four-particle 
tree amplitude corresponding to the interaction between them\footnote{It is perhaps worth emphasising that the mass $M$ appearing explicitly 
in this formula is inert under gravitational duality: the duality rotation acts on the amplitudes, not the momenta.}.

In fact, we do not need the full four-point amplitude, but only the coefficient of the $q^2$ pole. Terms in the amplitude without this pole (contact
terms) cannot contribute to the impulse, as reviewed in appendix~\ref{sec:integrals}. This fact was useful to us in the classical computations
of section~\ref{sec:classicalKTNimpulse}. In the context of amplitudes, the simplification is clear: we can reconstruct all relevant terms in
the amplitude by cutting only the $t$ channel pole, corresponding to the exchange of a photon.

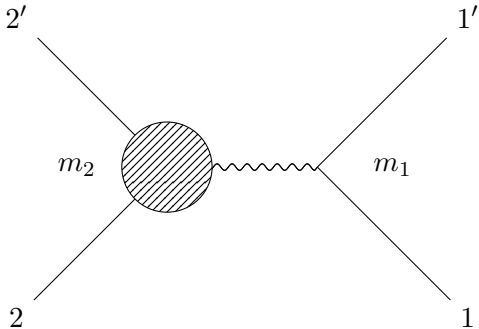
\begin{figure}[H]
	\centering
	\begin{tikzpicture}[scale=1]
	\begin{feynman}  
	\vertex (a) at (-4,2) {$2'$};
	\vertex (b) at (-4,-2) {$2$};
	\vertex (c) at (2,-2) {$1$};
	\vertex (d) at (2,2) {$1'$};
	\vertex (r) at (0,0);
	\vertex (l) at (-2,0) ;
	\diagram* {
		(a) -- [plain] (l) -- [photon] (r) -- [plain] (d),
		(b) -- [plain] (l) -- [photon] (r) -- [plain] (c),
	};
	\draw[preaction={fill, white},pattern=north east lines] (-2,0) ellipse (0.6cm and 0.6cm);
	\draw (-3.2,-0.25) node[above] {$m_2$};
	\draw (1,-0.25) node[above] {$m_1$};
	\end{feynman}
	\end{tikzpicture}
	\caption{Electromagnetic probe of a charged, spinning particle}
	\label{EMsetup}
\end{figure}

Let us now describe our construction of the relevant four-point amplitude, shown in 
figure~\ref{EMsetup}. We have a test scalar of mass $m_1$ and charge $e_1$, probing the electromagnetic field generated by 
a second particle which has (classical) spin $a^\mu$, mass $m_2$ and charge $e_2$.

Two three-point amplitudes are involved in this figure. One is the Coulomb three-point amplitude~\eqref{eq:CoulombAmplitude} for the test particle.
For the background, we take the amplitude for the spinning dyon, obtained by multiplying the Coulomb amplitude with both a 
duality phase~\eqref{eq:dualityPhase} and a Newman-Janis phase~\eqref{eq:NJphase}. Thus the amplitudes are
\begin{equation}\label{3pts2}
\cl{A}_3[1,1',q^{\pm}] = \sqrt{2}e_1M_1x_1^\pm,~~~~~\cl{A}_3[2^s,2'^s,q^{\pm}] = \sqrt{2}e_1M_2x_2^\pm e^{\pm(i\theta + q\cdot a)} \,.
\end{equation}
Amplitudes for a dyon or \rootKerr~background are obtained by setting appropriate parameters to zero. 

We ``glue'' these three-point amplitudes to construct the four-particle amplitude in the $q^2\rightarrow 0$ limit, in this case given by
\begin{equation}
\cl{A}_4 = \frac{2e_1e_2M_1M_2}{q^2}\left[\frac{x_1}{x_2}e^{-i\theta -q\cdot a} + \frac{x_2}{x_1}e^{i\theta+q\cdot a}\right].
\end{equation}

To compare to classical results from section~\ref{sec:impulse} it is helpful to rewrite the four-point amplitude in terms of the natural
variables of the classical theory.
Using the definition of the $x$ variables in eq.(\ref{xDef}), we can write\footnote{See Appendix B of \cite{Huang:2019cja} for a detailed derivation.}
\eq\label{Xratio}
\frac{x_1}{x_2}=-\frac{1}{M_1M_2}\frac{\langle \eta|p_1|q]}{\langle \eta q\rangle}\frac{\langle q|p_2|\eta]}{[q\eta]}=\frac{(p_1\cdot p_2)}{M_1M_2}{+}i\frac{\epsilon(\eta, p_1, q, p_2)}{M_1M_2(q\cdot \eta)} = u_1\cdot u_2 +i\frac{\epsilon(\eta, u_1, q, u_2)}{q\cdot\eta}\,,
\eqe
where $u_i \equiv p_i/M_i$ and $|\eta\rangle[\eta|$ is an auxiliary null vector which encodes the direction of the Dirac-string like singularity.
The fact that the amplitude depends on the position of the Dirac string has been discussed in~\cite{Caron-Huot:2018ape}: while the amplitude is expected to be gauge invariant, it can transform non-trivially under large gauge transformations. Since the string singularity remains visible at large distances, gauge transformations that do not vanish at infinity can move the position of the Dirac string, and thus the amplitude is expected to depend on its position. However, this dependence must drop out of physical observables such as the classical impulse, and indeed it does. To see this,  we plug 
our result~\eqref{Xratio} for the $x$-ratio into the amplitude, finding
\begin{equation}
\begin{aligned}
\cl{A}_4 &= \frac{2e_1 e_2M_1M_2}{q^2}\left[ u_1\cdot u_2 \left( e^{-q \cdot a -i \theta} + e^{q \cdot a + i \theta}  \right)
+ i\frac{\epsilon(\eta, u_1,q, u_2)}{q\cdot\eta} \left( e^{-q \cdot a -i \theta} - e^{q \cdot a + i \theta}  \right) \right] \,.
\end{aligned}
\end{equation}
With this amplitude in hand, equation~\eqref{impulseamp} directly gives us the classical impulse in the form
\begin{equation}
	\begin{aligned}
\Delta p^\mu_1 &= \frac{ie_1e_2}{2}\int \dd^4q \,\hat{\delta}(q\cdot u_1)\hat{\delta}(q\cdot u_2)e^{iq\cdot b} \frac{q^\mu}{q^2} \times \\
&\qquad\qquad\times\left[ u_1\cdot u_2 \left( e^{-q \cdot a -i \theta} + e^{q \cdot a + i \theta}  \right)
+ i\frac{\epsilon(\eta, u_1,q, u_2)}{q\cdot\eta} \left( e^{-q \cdot a -i \theta} - e^{q \cdot a + i \theta}  \right) \right] \,.
\label{eq:ampImpuseEMstep}
\end{aligned}
\end{equation}

We have not yet removed the apparent dependence on the unphysical vector $\eta$. It is not hard to do so: the basic idea was already
used in section~\ref{sec:classicalKTNimpulse} where we sometimes found it useful to use a dual basis. In the present context, we can resolve $q^\mu$ on to a dual basis given by $\epsilon^\mu(q, u_1, u_2)$, $\epsilon^\mu(u_1, u_2, \eta)$, $\epsilon^\mu(u_2, \eta, q)$, 
and $\epsilon^\mu(\eta, q, u_1)$. The details are discussed in appendix~\ref{sec:kinematics}. Taking advantage of 
equation~\eqref{eq:qReplacement} and the fact that polynomials in $q^2$ do not contribute, the impulse becomes
\[
\Delta p_1^\mu = \Re i e_1 e_2 \int \! \hat{\sd}^4 q \, \del(q \cdot u_1) \del(q \cdot u_2) e^{i q \cdot b} \frac{1}{q^2} &\left[ q^\mu \, u_1 \cdot u_2 - i \epsilon^\mu(q, u_1, u_2)  \right] e^{-(q \cdot a + i \theta)}  \,.
\]
This is precisely the expression~\eqref{EMimpulse2} we found by integrating the classical equations of motion.
Thus we have confirmed that the assignments of three-point amplitudes in figure~\ref{fig:amplitudeCube} are consistent with
our knowledge of the impulse in the classical backgrounds of figure~\ref{fig:solutionCube}.

\subsection{Gravitational Impulses from Amplitudes}

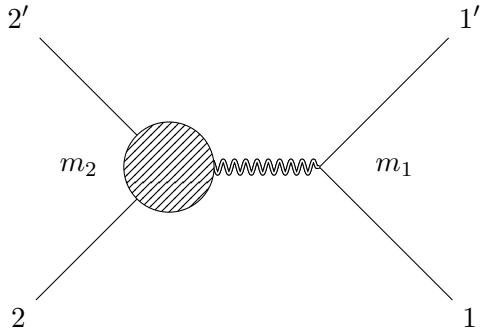
\begin{figure}[H]
	\centering
	\begin{tikzpicture}[scale=1]
	\begin{feynman}  
	\vertex (a) at (-4,2) {$2'$};
	\vertex (b) at (-4,-2) {$2$};
	\vertex (c) at (2,-2) {$1$};
	\vertex (d) at (2,2) {$1'$};
	\vertex (r) at (0,0);
	\vertex (l) at (-2,0) ;
	\diagram* {
		(a) -- [plain] (l) -- [graviton] (r) -- [plain] (d),
		(b) -- [plain] (l) -- [graviton] (r) -- [plain] (c),
	};
	\draw[preaction={fill, white},pattern=north east lines] (-2,0) ellipse (0.6cm and 0.6cm);
	\draw (-3.2,-0.25) node[above] {$m_2$};
	\draw (1,-0.25) node[above] {$m_1$};
	\end{feynman}
	\end{tikzpicture}
	\caption{Gravitational probe of a charged, spinning particle}
	\label{setup}
\end{figure}

We now turn to the computation of the impulse in gravity using amplitudes. We scatter a probe mass off (linearised) Kerr-Taub-NUT,
obtaining the Kerr and Taub-NUT cases as limits. 
We will again consider a scalar test-particle with mass $m_1$, probing a spacetime generated by particle 2, which has classical spin $a^\mu$ and mass $m_2$, as in Fig. \ref{setup}. Our aim here is to calculate the impulse imparted on the test particle due to its interaction with the spacetime generated by particle 2 at first-order (1PM) of the post-Minkowskian expansion of the metric.

We will proceed as in the electromagnetic case and evaluate the impulse by computing the on-shell four-particle amplitudes arising from Fig. \ref{setup}. In particular, we highlight the efficiency afforded by modern scattering amplitude techniques in calculating observable quantities in the PM expansion.

The long range behavior of black holes can be well captured by minimally coupled particles \cite{Guevara:2018wpp,Chung:2018kqs,Huang:2019cja,Chung:2019duq,Guevara:2019fsj,Moynihan:2019bor,Chung:2019yfs}, a fact which can be attributed to the no hair theorem, where the classical solutions are labelled by the same set of quantum numbers as elementary particles: mass, charge and spin. Although not strictly a black hole solution, the Kerr-Taub-NUT metric can also be described by a similar set of quantum numbers, in this case the mass, NUT charge (dual mass), and spin. 

Proceeding as we did in the electromagnetic case, the relevant three-particle amplitudes for this scattering process at lowest order are\footnote{By
choice of convention, we could introduce a factor 2 in the duality and Newman-Janis phases as in figure~\ref{fig:amplitudeCube}. Such a choice
would have the virtue of making the double copy more evident. However, we chose not to do so for notational convenience.}
\begin{equation}\label{3pts3}
\cl{M}_3[1,1',q^{\pm2}] = \frac{\kappa}{2}(M_1x_{1}^\pm)^2,~~~~~\cl{M}_3[2^s,2'^s,q^{\pm 2}] = \frac{\kappa}{2}M^2_2 x_{2}^{\pm2}e^{\pm (i\theta+q\cdot a)}\,,
\end{equation}
where $\kappa^2 = 32 \pi G$.
Note that the phase picks up an extra sign with the change of helicity. This extra sign can be attributed to the fact that the positive (negative) helicity state correspond to the (anti)-self dual Riemann tensor in the linear approximation. 
We have defined the duality and NJ phases with a factor of two with respect to the electromagnetic case in order to highlight the double copy structure. However, this is purely conventional and could be absorbed into a re-definition of the phases.

It is a straightforward matter to glue these into a four-point amplitude in the $q^2\rightarrow 0$ limit. Let us first discuss the Taub-NUT
case where $a = 0$, so that
\eqa
\cl{M}_4&=&\left(\frac{\kappa}{2}\right)^2\frac{M^2_1M^2_2}{q^2}\left(\left(\frac{x_1}{x_2}\right)^2e^{i\theta } + \left(\frac{x_2}{x_1}\right)^2e^{- i\theta}\right)\nonumber\\
&=&\left(\frac{\kappa}{2}\right)^2\frac{M^2_1M^2_2}{q^2}\left(2\cos \theta \left(\frac{(p_1\cdot p_2)^2}{M^2_1M^2_2}{-}\frac{\epsilon(\eta, p_1, q, p_2)^2}{M_1^2M_2^2(q\cdot \eta)^2} \right) {-}4\sin \theta\frac{(p_1\cdot p_2)\epsilon(\eta, p_1, q, p_2)}{M^2_1M^2_2(q\cdot \eta)} \right)\nonumber\\
&=&\kappa^2\frac{M^2_1M^2_2}{q^2}\left(\cos \theta \left(\frac{(p_1\cdot p_2)^2}{M^2_1M^2_2}{-}\frac{1}{2} \right) {-}\sin \theta\frac{(p_1\cdot p_2)\epsilon(\eta, p_1, q, p_2)}{M^2_1M^2_2(q\cdot \eta)} \right)\,.\nonumber\\
\eqae
We used the result~\eqref{Xratio} to simplify the ratio of $x$ factors, and removed the squared determinant with the help of the kinematics
discussed in appendix~\ref{sec:kinematics}, specifically equation~\eqref{eq:detSquare}.
Notice that the dependence on the Misner string singularity \cite{Misner:1963fr,Misner:1965zz}, parameterised by $\eta$, drops out when the phase $\theta$ is zero, i.e. in the Schwarzschild case. This is as expected: when the magnetic mass vanishes, the amplitude is gauge invariant.

Next, we consider $a \neq 0$. The amplitude (ignoring classically irrelevant contact terms) is
\begin{align}\label{KTNSol}
\cl{M}_4 &= \left(\frac{\kappa}{2}\right)^2\frac{M^2_1M^2_2}{q^2}\left(\left(\frac{x_1}{x_2}\right)^2e^{i\theta+q\cdot a} + \left(\frac{x_2}{x_1}\right)^2e^{- (i\theta+q\cdot a)}\right).
\end{align}
We can again simplify the $x$ ratio using equation~\eqref{Xratio}, and find the impulse on the probe using~\eqref{impulseamp}. The result is
\begin{equation}
\begin{aligned}
\Delta p^\mu_1 &=  \left(\frac{\kappa}{2}\right)^2\frac{M_1M_2}{4}\int \dd^4q\,\hat{\delta}(q{\cdot }u_1)\hat{\delta}(q{\cdot} u_2)e^{iq{\cdot} b}\, \frac{iq^\mu}{q^2}
\\& \qquad \qquad\qquad\qquad\qquad \times
\sum_{\pm}\left(\cosh 2w {\mp}2 i \cosh w \frac{\epsilon(\eta, u_1, q, u_2)}{q\cdot \eta}\right)e^{ \pm (i\theta{+}q\cdot a)} \,.
\end{aligned}
\end{equation}
where we simplified the product $\epsilon(\eta, u_1, q, u_2)^2$ using equation~\eqref{eq:detSquare}.

As in the electromagnetic case, the vector $\eta$ must drop out irrespective of the value of $\theta$ since the impulse is an observable. The 
details are precisely as in the electromagnetic case, allowing us again 
to make the replacement~\eqref{eq:qReplacement}. We find
\begin{equation}
\begin{aligned}
\Delta p^\mu_1 &=  \Re 4 \pi G {M_1M_2}\int \dd^4q\,\hat{\delta}(q{\cdot }u_1)\hat{\delta}(q{\cdot} u_2) \frac{e^{iq{\cdot} b}}{q^2}
\\& \qquad \qquad\qquad\qquad  \qquad\times 
\left(iq^\mu\cosh 2w - 2  \epsilon^{\mu}(u_1,u_2,q)\cosh w \right)e^{ -(q{\cdot} a+ i\theta)}.
\end{aligned}
\end{equation}
which coincides with \eqref{gravityImpulse}.

We have a detailed match: impulses computed using the double copy of amplitudes are equal to impulses computed in classical
backgrounds which we claim should double copy. 

It is amusing to demonstrate the power if the scattering amplitudes approach by writing down the impulse on two Kerr-Taub-NUTs scattering
at leading order. We need only introduce separate duality and NJ phases for each object; otherwise the computation is precisely as above. The
result is, by inspection,
\[
\Delta p^\mu_1 &=  \Re 4 \pi G {M_1M_2}\int \dd^4q\,\hat{\delta}(q{\cdot }u_1)\hat{\delta}(q{\cdot} u_2) \frac{e^{iq{\cdot} b}}{q^2}
\\& \qquad \qquad\qquad\qquad  \qquad\times 
\left(iq^\mu\cosh 2w - 2  \epsilon^{\mu}(u_1,u_2,q)\cosh w \right)e^{- i(\theta_2 - \theta_1){-}q{\cdot} (a_2 + a_1)} \,,
\]
where the duality angles and spins of particles 1 and 2 are $\theta_1$, $\theta_2$ and $a_1$, $a_2$. The $q$ integral has been 
discussed elsewhere~\cite{Kosower:2018adc},
and we provide the explicit evaluation in equation~\eqref{eq:integralEvaluation}. Note that the duality angles subtract 
because of the relative signs of the duality phases in the three-point amplitudes~\eqref{eq:NJphase}. The spins add, because the momentum $q$ is
incoming from the point of view of one three-point amplitude, but outgoing for the other. It is not so trivial to reproduce this computation using
traditional classical methods known to us.

We could also easily compute the change in spin (``angular
impulse'') on these objects during scattering using the methods of~\cite{Maybee:2019jus,Guevara:2019fsj}, which require only the scattering amplitude.

\section{Discussion}\label{sec:discussion}

The key idea in our work was to use duality, the Newman-Janis shift, and the double copy as guides through a network of solutions and 
an associated network of scattering amplitudes. 

Beginning from the EM field of a Coulomb charge, these operations build up the class of eight solutions shown in
figure~\ref{fig:solutionCube}. Crucially, the same three operations also act on three-point amplitudes by a simple multiplication 
of phase factors. We were therefore able to associate a unique scattering amplitude to each of our solutions, beginning with the well-known 
amplitude for a photon interacting with a heavy point charge. The result is the network of amplitudes shown in 
figure~\ref{fig:amplitudeCube}.
This unambiguous connection between amplitude and solution is one of the principal achievements of our work. 

Our knowledge of which amplitude is associated to which solution provides strong evidence in support of the classical double copy,
especially in its formulation through the Weyl spinor~\cite{Luna:2018dpt}. In spite of considerable work, the classical double copy is 
still based on 
matching symmetries and structural properties of EM and GR solutions.
From this perspective it is very satisfying that we have been able to link amplitude to solution so directly.
We went further by explicitly showing that a particular classical observable --- the impulse --- maps from gauge to gravitational solution 
as predicted by the double copy. We achieved this by computing the impulse by two entirely different methods. The first
was to use the known classical solutions and equations of motion, thereby determining the impulse due to an EM background
and due to its classical double copy. The second method made use of amplitudes, and of the double copy for amplitudes.
Building on the three-point amplitude associated with each of our solutions,
we constructed a four-point amplitude describing the leading order (tree) interaction between a probe particle and the background.
These four-point amplitudes double copy as expected. The impulse is computable from four point amplitudes, and we verified agreement
with the previous classical computation. In this way we matched impulses related by the classical double copy to impulses related
by the amplitudes double copy.
The match between both computations provides firm evidence supporting the Weyl double copy.

Another set of physical quantities which transform simply under our three operations are the asymptotic charges. We have seen throughout
this work that the action of our operations on asymptotic charges can be understood clearly when the charges are formulated using
Newman-Penrose scalars. The Newman-Penrose formalism is also closely connected to the classical double copy~\cite{Luna:2018dpt,Elor:2020nqe}.
Moreover, certain scattering amplitudes are closely associated with the NP scalars, or rather with the Maxwell and Weyl spinors which are in one-to-one correspondence
with them~\cite{ampTalk}. It would be exciting to explore the mutual connections among these topics in more depth.

At the level of classical solutions, one novel aspect of our work is that we have not expanded in the spin parameter $a$. This contrasts
with previous work on Kerr-Taub-NUT, for instance. It is important to note that if one sets the mass $m$ and the NUT charge $\ell$ of
Kerr-Taub-NUT to zero, the curvature vanishes for \emph{any} value of the spin. But the curvature is non-vanishing for other (real) values
of $m$ or $\ell$. Thus the first post-Minkowskian expansion forces us to linearise in $m$ and $\ell$, but has nothing to say about the spin.
The issue of whether a singularity is naked or shrouded by a horizon is not yet accessible to scattering amplitudes, which in any case
require a perturbative expansion, so censorship considerations need not force us to assume that $a$ is small when we expand in $m$ and $\ell$.

Our work suggests a number of interesting questions for future work. For example, there are several proposals for how to implement the 
classical double copy to all orders~\cite{Monteiro:2014cda,Luna:2015paa,Luna:2018dpt,Elor:2020nqe}. These proposals differ in their
details, and apply to different sets 
of solutions. However they agree when their domains of validity overlap. Our method of testing a proposed classical double copy 
using the impulse may be useful for excluding or refining proposals.

Beyond linearised order, we must also face the fact that the double copy relates Yang-Mills theory (not electromagnetism) to gravity. 
Our work highlights a non-trivial test for the classical double copy: we have checked that the leading order impulse double copies as 
expected, but what about the next-to-leading order correction? The correction could be computed both by standard classical methods,
and also using higher order amplitudes. If these computations match (as we expect) it will provide further evidence that
the classical double copy holds to all orders. Curiously, a recent investigation~\cite{delaCruz:2020bbn} of the impulse at next-to-leading
 order in Yang-Mills theory
found that the non-abelian terms in the next-to-leading order amplitude did not contribute: it would be interesting to understand how this
can be consistent with the non-linearity of classical general relativity at this order.

We benefited from an understanding of gravitational duality at leading order which, for example, allowed us to identify the
three-point amplitude associated with Kerr-Taub-NUT.
It would be very interesting to understand gravitational duality beyond linearised order in more detail. The double-Kerr-Schild form
of the Kerr-Taub-NUT metric, which exists in a complexified spacetime or a different signature, is a promising place to start~\cite{Luna:2018dpt}.

From the traditional perspective of scattering amplitudes, it seems quite amazing that a simple amplitude exists for an esoteric-seeming object
like Kerr-Taub-NUT, which has a magnetic mass and a large classical spin. These properties are foreign to the particle physics objects
(quarks, gluons, leptons, \ldots) more commonly associated with amplitudes. We built on the explosion~\cite{Luna:2016due,Damour:2016gwp,Goldberger:2016iau,Cachazo:2017jef,Damour:2017zjx,Luna:2017dtq,Laddha:2018rle,Laddha:2018myi,Laddha:2018vbn,Bjerrum-Bohr:2018xdl,Plefka:2018dpa,Cheung:2018wkq,Sahoo:2018lxl,Kosower:2018adc,Bern:2019nnu,KoemansCollado:2019ggb,Brandhuber:2019qpg,Cristofoli:2019neg,Plefka:2019hmz,Laddha:2019yaj,Bern:2019crd,Damgaard:2019lfh,Kalin:2019rwq,Bjerrum-Bohr:2019kec,Kalin:2019inp,Huber:2019ugz,Saha:2019tub,Aoude:2020onz,Bern:2020gjj,Cheung:2020gyp,Cristofoli:2020uzm,Bern:2020buy,Parra-Martinez:2020dzs,Haddad:2020tvs,AccettulliHuber:2020oou,Cheung:2020sdj,A:2020lub,Kalin:2020fhe,Haddad:2020que,Kalin:2020lmz,Sahoo:2020ryf,DiVecchia:2020ymx,Guevara:2018wpp,Chung:2018kqs,Huang:2019cja,Chung:2019duq,Guevara:2019fsj,Moynihan:2019bor,Chung:2019yfs,Mogull:2020sak} of interest in applications of
scattering amplitudes to gravitational wave physics, where the spinning Kerr solution is particularly important.
Magnetic charges are also unusual
in the context of amplitudes because charge quantisation indicates that the product of electric and magnetic charge is a poor expansion
parameter. However, applications of amplitudes to classical physics involve an expansion in scattering angles~\cite{Kosower:2018adc} rather than in charges,
so that a perturbative series exists even in the presence of magnetic charge. Perhaps scattering amplitudes will find a still wider 
application beyond particle colliders.

\section*{Acknowledgements}

WTE is supported by an STFC consolidated grant, under grant no. ST/P000703/1, and also by the Czech Science Foundation GA\v{C}R, 
project 20-16531Y.
Meanwhile, NM and DOC are supported by the STFC grant ST/P0000630/1. NM is also supported by funding from the DST/NRF SARChI in 
Physical Cosmology. Any opinion, finding and conclusion or recommendation expressed in this material is that of the authors and the 
NRF does not accept any liability in this regard.

Some of our figures were produced with the help of TikZ-Feynman~\cite{Ellis:2016jkw}.

\appendix

\section{Conventions}
\label{sec:conventions}

We work in signature $(-,+,+,+)$ throughout. For the Levi-Civita tensors, we set
\[
\epsilon_{0123} &= +1 \,, \\
\epsilon_{123} &= +1 \,.
\]
As we frequently encounter expressions (anti-)symmetrised over pairs of indices, we define
\[
x_{[\mu} y_{\nu]} &= \frac12 (x_\mu y_\nu - x_\nu y_\mu) \,, \\
x_{(\mu} y_{\nu)} &= \frac12 (x_\mu y_\nu + x_\nu y_\mu) \,.
\]
We frequently Fourier transform to momentum space, writing
\[
f(x) &= \int \dd^4 q \, e^{i q\cdot x} \,\tilde f(q) \, \\
\tilde f(q) &= \int \sd^4 x \, e^{-i q\cdot x} \, f(x) \,.
\]
Hatted measures and delta functions absorb factors of $2\pi$:
\[
\dd q &= \frac{1}{2\pi} \sd q\,, \\
\del (q) &= 2\pi \delta(q) \,.
\]
The Levi-Civita tensor plays an important role in our work. We introduce the notation
\[
\label{eq:epsContractionDef}
\epsilon^\mu(a,b,c) &= \ep^{\mu\nu\rho\sigma} a_{\mu}b_{\nu}c_{\sigma} \,, \\
\epsilon^{\mu\nu}(a, b) &= \epsilon^{\mu\nu\rho\sigma} a_\rho b_\sigma \,, \\
\epsilon^{\mu\nu\rho}(a) &=\epsilon^{\mu\nu\rho\sigma} a_\sigma \,,  \\
\]
to clarify these contractions.

We write the Clifford algebra as
\[
\sigma^a \tilde \sigma^b - \sigma^b \tilde \sigma^a = -2 \eta^{ab} \,,
\]
where $\eta$ is the Minkowski inner product on tangent spaces, and $a, b$ are frame indices. We pick
\[
\sigma^a &= (1, \boldsymbol{\sigma}) \,, \\
\tilde \sigma^a &= (1, - \boldsymbol{\sigma}) \,.
\]
Furthermore, we write
\[
\sigma^{ab} = \frac14 (\sigma^a \tilde \sigma^b - \sigma^b \tilde \sigma^a) \,.
\]
We raise and lower spinor indices with the two-dimensional Levi-Civita $\epsilon$. Aside from the sign of the metric in the Clifford algebra, we adopt the detailed spinor-helicity conventions described in Appendix A of~\cite{Chung:2018kqs}.

Turning to electrodynamics, the Maxwell equations and Lorentz force are
\[
\partial_\mu F^{\mu\nu}(x) &= - J^\nu(x) \,, \\
\frac{\sd p^\mu}{\sd\tau} &= e F^{\mu\nu}(r(\tau)) \,u_\nu \,,
\]
where $e$ is the electric charge of a particle with worldline $r(\tau)$. We define (as usual)
\[
F_{\mu\nu} = \partial_\mu A_\nu - \partial_\nu A_\mu \,,
\]
so that the electric and magnetic fields are
\[
\label{eq:ebFieldEmbedding}
E_i &= F_{i0} \,, \\
B_i &= \frac12 \epsilon_{ijk} F_{jk} \,.
\]
In particular, the scalar electric potential is $A^0$. The electric and magnetic fields can also be written in terms of the dual tensor
\[
\tilde F_{\mu\nu} = \frac12 \epsilon_{\mu\nu\rho\sigma} F^{\rho\sigma}\,.
\]
In our conventions, the explicit relation is
\[
E_i &= \frac12 \epsilon_{ijk} \tilde F_{jk} \,\\
B_i  &= - \tilde F_{i0} \,.
\]

The geodesic equation is
\[
\frac{\sd u^\mu}{\sd\tau} = - \Gamma^\mu{}_{\nu\rho} u^\nu u^\rho \,.
\]
We define the Riemann curvature to be
\[
R^\mu{}_{\nu\rho\sigma} = \partial_\rho \Gamma^\mu{}_{\nu\sigma} - \partial_\sigma \Gamma^\mu{}_{\nu\rho} + \Gamma^\mu{}_{\rho\lambda} \Gamma^\lambda{}_{\nu\sigma} - \Gamma^\mu{}_{\sigma\lambda} \Gamma^\lambda{}_{\nu\rho} \,.
\]

\section{Duality, Kerr and Spinning NUTs}
\label{app:duality}

In the main text, we used the fact that a duality rotation deforms the linearised Kerr metric into the linearised Kerr-Taub-NUT metric. This duality
is evident from the Weyl spinor given in equation~\eqref{eq:linPsi2}. This spinor determines the Weyl tensor, which equals the Riemann tensor
in our vacuum application. So it must be the case that the Riemann tensor of a pure spinning NUT is the dual of the Riemann tensor of Kerr, once
the masses and NUT charges are interchanged.

We can also demonstrate that the duality holds directly at the level of the Riemann tensors, which may be a more familiar language. The price
is that the expressions are lengthier. In this appendix, we explain how to see the duality in the simpler case when the spin $a = 0$. In other
words, we show that the curvature of the static NUT contribution to the linearised Taub-NUT metric is equal to the dual of the linearised Schwarzschild
curvature.

We attach a mathematica notebook performing precisely the same calculation for the $a \neq 0$ case, with the final result that the Riemann
curvature of the spinning NUT is equal to the dual of the Riemann curvature of Kerr as expected.

To keep the notation straight, it is useful to write the linearised curvature in a tensorial form
\[
\mathcal{R} := \frac{1}{4} R_{\mu\nu\rho\sigma} (\sd x^\mu \wedge \sd x^\nu) \otimes (\sd x^\rho \wedge \sd x^\sigma) \,.
\]
We will work (in this appendix) in spherical coordinates, using the frame
\[
\label{eq:sphericalFrame}
e^t= \sd t\,, \quad e^r = \sd r\,, \quad e^\theta = r \, \sd\theta\,, \quad e^\phi = r \sin \theta \, \sd \phi \,.
\]
Then we may also write the curvature as
\[
\mathcal{R} = \frac14 R_{abcd} (e^a \wedge e^b) \otimes (e^c \wedge e^d) \,.
\]
For example, the explicit linearised curvature of Schwarzschild is
\[
\mathcal{R}^\textrm{Schw} = \frac{1}{r^3} &\left[-2 e^r \wedge e^t \otimes e^r \wedge e^t + 2 e^\theta \wedge e^\phi \otimes e^\theta \wedge e^\phi + e^\theta \wedge e^t \otimes e^\theta \wedge e^t  \right.
\\
& \left.
\quad + e^\phi \wedge e^t \otimes e^\phi \wedge e^t - e^r \wedge e^\theta \otimes e^r \wedge e^\theta - e^r \wedge e^\phi \otimes e^r \wedge e^\phi
\right] \,.
\]
It is easy to compute the dual curvature in this form. The Hodge dual is
\[
* e^a \wedge e^b = \frac12 \epsilon^{ab}{}_{cd} \, e^c \wedge e^d 
\]
as usual. The duality \eqref{dual} acts only on the left two-form in our tensorial curvature
\[
* \mathcal{R} &\equiv \left(\frac12 \epsilon_{ab}{}^{ef} R_{efcd} \right)  \frac 14 (e^a \wedge e^b) \otimes (e^c \wedge e^d) \\
&= \frac 14 R_{abcd} (* e^a \wedge e^b) \otimes (e^c \wedge e^d) \,.
\]

The linearised curvature of Taub-NUT is the sum of the linearised Schwarzschild curvature, and the curvature due to the NUT. This second 
contribution is, explicitly,
\[
\mathcal{R}^\textrm{NUT} = -\frac{\ell}{r^3} &\left[-2 e^\theta \wedge e^\phi \otimes e^r \wedge e^t + 2 e^t \wedge e^r \otimes e^\theta \wedge e^\phi + e^\phi \wedge e^r \otimes e^\theta \wedge e^t  \right.
\\
& \left.
\quad + e^r \wedge e^\theta \otimes e^\phi \wedge e^t - e^t \wedge e^\phi \otimes e^r \wedge e^\theta - e^\theta \wedge e^t \otimes e^r \wedge e^\phi
\right] \,.
\]
It is easy to check that
\[
\frac{1}{m} * \mathcal{R}^\textrm{Schw} = \frac{1}{\ell} \mathcal{R}^\textrm{NUT} \,.
\]
Consequently, we may write the linearised Riemann tensor of Taub-NUT in the form
\[
R^\textrm{TN}_{\mu\nu\rho\sigma} = GM \Re e^{-i \theta}  \left(\Theta^\textrm{Schw}_{\mu\nu\rho\sigma} + i \tilde \Theta^\textrm{Schw}_{\mu\nu\rho\sigma} \right) \,,
\]
where $m = GM \cos \theta$ and $\ell = GM \sin \theta$, and (as in the main text)  the $\Theta$ tensors extract the linear dependence of the curvature tensors on the mass: 
\[
\Theta^\textrm{Schw}_{\mu\nu\rho\sigma} = \frac{1}{m} R^\textrm{Schw}_{\mu\nu\rho\sigma} \,.
\]

For the $a\neq 0$ case, we use spheroidal coordinates and the spheroidal frame generalising the spherical one~\eqref{eq:sphericalFrame} above. 
The calculation is otherwise exactly analogous, and we find 
\[
\frac{1}{m} * \mathcal{R}^\textrm{Kerr} = \frac{1}{\ell} \mathcal{R}^\textrm{spinning-NUT} \,.
\]
Explicit details are contained in the notebook.

It follows that the linearised Riemann tensor for Kerr-Taub-NUT can be written as
\[
R^\textrm{KTN}_{\mu\nu\rho\sigma} &= m \, \Theta^\textrm{Kerr}_{\mu\nu\rho\sigma} + \ell \, \Theta^\textrm{spinning-NUT}_{\mu\nu\rho\sigma} \\
&= GM \Re e^{-i \theta} \left(\Theta^\textrm{Kerr}_{\mu\nu\rho\sigma} + i \tilde \Theta^\textrm{Kerr}_{\mu\nu\rho\sigma} \right) \,,
\]
where, again, the parameters are related by $m = GM \cos \theta$ and $\ell = GM \sin \theta$.

\section{Scattering Kinematics}
\label{sec:kinematics}

In the main text of the paper, it was important to understand the scattering of two classical objects in detail. We gather some aspects of the
kinematics in this appendix.

The unperturbed trajectories of our objects are the straight lines
\[
x_1(\tau) = b + u_1 \tau \,,
x_2(\tau) = u_2 \tau \,.
\]
The impact parameter is constrained to satisfy 
\[
u_1 \cdot b = 0 = u_2 \cdot b \,.
\]
In either the classical or quantum treatments of the scattering, the total change in momentum (the impulse) arises as an integration over an
object $q^\mu$. Classically, this $q$ is a wavenumber. Quantum mechanically, it is the momentum of the photons or gravitons mediating the
interaction. The amplitude, which is large in the classical approximation, weights the momentum exchanged by the individual messenger bosons.
In any case, the vector $q$ satisfies the constraints
\[
u_1 \cdot q = 0 = u_2 \cdot q \,.
\]
Our classical and quantum computations involved additional vectors: the spin $a$ and (in the amplitudes approach) a gauge $\eta$. Thus we
have a basis of vectors: $u_1$, $u_2$, $q$ and $\eta$. (The impact parameter could be included in a basis, though it appears in the computations
in a rather simple way and needs no further manipulation. We may replace $\eta$ with $a$ below where appropriate.)

In the main text, it was often useful to resolve these vectors on a dual basis. For example, consider
\[
q^\mu = A \, \epsilon^\mu(q, u_1, u_2) + B \, \epsilon^\mu(u_1, u_2, \eta) + C \, \epsilon^\mu(u_2, \eta, q) + D\, \epsilon^\mu(\eta, q, u_1) \,,
\]
where $A$, $B$, $C$ and $D$ are coefficients to be determined. But since $q \cdot u_1 = 0 = q \cdot u_2$ on the support of the integral, it 
immediately follows that $C = 0 = D$. Continuing, we find
\[
q^\mu \, \epsilon(\eta, q, u_1, u_2) = q \cdot \eta \, \epsilon^\mu(q, u_1, u_2) - q^2 \, \epsilon^\mu(u_1, u_2, \eta) \,.
\]
Now, the impulse computed in the main text has the property that polynomials in $q^2$ do not contribute. Thus, we may replace
\[
q^\mu \frac{\epsilon(\eta, u_1,q, u_2)}{q\cdot\eta} \rightarrow  - \epsilon^\mu(q, u_1, u_2)
\label{eq:qReplacement}
\]
in the impulse.

We also encounter the determinant $\epsilon(\eta, q, u_1, u_2)$ at various points. For simplicity, we assume that $\eta \cdot u_2 = 0$; this would
be true by definition if $\eta = a$ and is a valid gauge choice for the $\eta$ encountered is scattering amplitudes. 
First, it is straightforward to compute the square of this determinant which is
\[
\epsilon(\eta, q, u_1, u_2)^2 &= - \det \begin{pmatrix}
\eta^2 & \eta \cdot q & \eta \cdot u_1 & \eta \cdot u_2 \\
\eta \cdot q & q^2 & 0 & 0 \\
\eta \cdot u_1 & 0 & -1 & u_1 \cdot u_2 \\
\eta \cdot u_2 & 0 & u_1 \cdot u_2 & -1
\end{pmatrix} \\
& = - ( \cosh^2 w - 1) (\eta \cdot q)^2 + q^2 ( (\eta \cdot u_1)^2 - (\cosh^2 w -1) \eta^2) \,,
\]
where we define the rapidity $w$ via
\[
\cosh w = - u_1 \cdot u_2 \,.
\]
Thus, when we can neglect $q^2$, 
\[
\label{eq:detSquare}
\epsilon(\eta, q, u_1, u_2)^2 = - ( \cosh^2 w - 1) (\eta \cdot q)^2 \,.
\]

\section{Fourier integrals}
\label{sec:integrals}

Finally, we provide a few remarks about the Fourier integrals appearing in computations of the impulse. Whether in electromagnetism or gravity,
working classically or using amplitudes, the impulse has the structure
\[
\Delta p^\mu = \mathcal{N} \int \dd^4 q \, \del(q\cdot u_1) \del(q \cdot u_2) \, e^{i q\cdot b} \frac{iq^\mu}{q^2} f(q) \,,
\]
where the function $f(q)$ is a scalar function, depending on the vector $q^\mu$ and perhaps other parameters (for example, the spin $a$). The number $\mathcal{N}$ here is a normalisation factor, and $u_1$ and $u_2$ are vectors having the interpretation of the particles' proper velocities
in the main text. Meanwhile, $b$ is the impact parameter vector, satisfying $u_i \cdot b = 0$ for $i = 1,2$.

It is important to realise that we may ignore terms in $f(q)$ which are polynomials in $q^2$. To see this, suppose the function has the structure
\[
f(q) = c_1 + c_2 \, q^2 \,,
\]
where $c_2$ has no dependence on $q$. It is easy to see that the contribution of $c_2$ to the impulse is proportional to $\delta^2(\vector{b})$, the
delta function in the two-dimensional space transverse to $u_1$ and $u_2$. Since $u_1 \cdot b = 0 = u_2 \cdot b$, the impact parameter is
exactly zero on the support of the delta function: the particles must collide head-on. This region is outside the domain of validity of our calculation,
which relies on small scattering angles and perturbative interactions. It is also outside the domain of validity of the 
classical approximation to the quantum field theory~\cite{Kosower:2018adc}. Similar comments apply to higher powers of $q^2$ in $f(q^2)$, 
which evaluate to derivatives of delta functions.
As a result, we may ignore polynomials of $q^2$ in $f(q)$, which is useful at various stages in the computation of the impulse.

Now let us provide the evaluation of the specific Fourier integral which appears in the impulse. We define
\[
\mathcal{I}^\mu \equiv \int \hat{\sd}^4 q \, \del(q\cdot u_1) \del(q \cdot u_2) \, e^{i q\cdot b} \frac{iq^\mu}{q^2} \,.
\]
The integral is easy to compute, and has been discussed in detail for example in reference~\cite{Kosower:2018adc}. The integral evaluates to
\[
\mathcal{I}^\mu = - \frac{1}{2\pi} \frac{1}{b^2 |\sinh w|} b^\mu \,.
\]
More generally, with a complex parameter, the integral is
\[
\label{eq:integralEvaluation}
\int \hat{\sd}^4 q \, \del(q\cdot u_1) \del(q \cdot u_2) \, e^{i q\cdot (b + i a)} \frac{iq^\mu}{q^2} = - \frac{1}{2\pi} \frac{1}{|\sinh w|} \frac{(b + i \Pi a)^\mu}{(b+i \Pi a)^2} \,,
\] 
where $\Pi^\mu{}_\nu$ is the projector onto the plane perpendicular to $u_1$ and $u_2$. Notice that only the components of $a$ in this plane are
present on the left-hand side of equation~\eqref{eq:integralEvaluation}; this is the origin of the projector. We can conveniently write the 
projector as
\[
\Pi^\mu{}_\nu = \frac{\epsilon^{\mu\alpha}(u_1, u_2) \epsilon_{\nu\alpha}(u_1, u_2)}{\cosh(w) - 1} \,.
\]

\bibliographystyle{JHEP}
\providecommand{\href}[2]{#2}\begingroup\raggedright\endgroup

\end{document}